\documentclass[a4paper,fleqn,usenatbib]{mnras}
\usepackage{newtxtext,newtxmath}
\usepackage[T1]{fontenc}
\usepackage{ae,aecompl}
\usepackage{graphicx}
\usepackage{threeparttable}

\newcommand{\etal}{\it et al. \rm }

\title[]{The Mass-to-light Ratios and the Star Formation Histories of Disk Galaxies}

\author[]{James Schombert$^{1}$\thanks{Contact e-mail:
\href{mailto:jschombe@uoregon.edu}{jschombe@uoregon.edu}},Stacy McGaugh$^{2}$ and Federico Lelli$^{3}$\thanks{ESO
Fellow}
\\
$^{1}$Department of Physics, University of Oregon, Eugene, OR 97403 \\
$^{2}$Department of Astronomy, Case Western Reserve University, Cleveland, OH 44106 \\
$^{3}$European Southern Observatory, Karl-Schwarschild-Strasse 2, Garching bei Munchen, Germany}

\date{Accepted 2018 Nov 26. Received 2018 Nov 5; in original form 2018 Aug 27}
\pubyear{2018}
\begin{document}
\label{firstpage}
\pagerange{\pageref{firstpage}--\pageref{lastpage}}
\maketitle

\begin{abstract}

\noindent We combine new data from the main sequence ($M_*$ versus SFR) of
star-forming galaxies and galaxy colors (from $GALEX$ to $Spitzer$) with a flexible
stellar population scheme to deduce the mass-to-light ratio ($\Upsilon_*$) of
star-forming galaxies from the SPARC and S$^4$G samples.  We find that the main
sequence for galaxies, particular the low-mass end, combined with the locus of galaxy
colors, constrains the possible star formation histories of disk and dwarf galaxies
to a similar shape found by Speagle \etal (2014).  Combining the deduced star
formation history with stellar population models in the literature produces reliable
$\Upsilon_*$ values as a function of galaxy color with an uncertainty of only 0.05
dex.  We provide prescriptions to deduce $\Upsilon_*$ for optical and near-IR
bandpasses, with near-IR bandpasses having the least uncertainty ($\Upsilon_*$ from
0.40 to 0.55).  We also provide the community with a webtool, with flexible stellar
population parameters, to generate their own $\Upsilon_*$ values over the wavelength
range for most galaxy surveys.

\end{abstract}

\begin{keywords}
techniques: photometric -- galaxies: star formation -- galaxies: stellar content
\end{keywords}

\section{Introduction}

With the discovery of the radial acceleration relation (McGaugh, Lelli \& Schombert
2016), it has become increasingly obvious that, on galactic scales, baryons play a
dominant role in the formation and evolution of galaxies.  The baryonic component of
galaxies is primarily gas (atomic, molecular and ionized) and stars (visible and
remnants).  The gaseous component in galaxies is heavily dominated by atomic gas
(e.g, Cortese, Catinella \& Janowiecki 2017), which is well measured using HI
observations and corrected by a factor of 1.33 to account for He and molecular
components.  The stellar baryon component is estimated from determination of the
total luminosity of a galaxy (at different wavelengths), then converted into a total
stellar mass value through multiplication of a mass to light ratio ($\Upsilon_*$), a
value deduced through some knowledge of the star formation history of the galaxy.  

In addition to probing the star formation history of a galaxy, a detailed knowledge
of the $\Upsilon_*$ is also a critical test for exotic theories such as MOND
(MOdified Newtonian Dynamics) and emergent gravity.  MOND proposes that the equations
of motion become scale-invariant at accelerations smaller than a characteristic
acceleration scale ($a_o = cH_o/6$, Milgrom 2009).  It predicts a correlation between
the observed centripetal acceleration and the acceleration from baryons (for a
typical rotating galaxy, this is due to gas and stars, thus the need for $\Upsilon_*$
to convert luminosity into stellar mass).  Emergent gravity proposes that gravity is
not a fundamental force but emerges from an underlying microscopic theory (Verlinde
2016).  It requires particular values of $\Upsilon_*$ in order to accommodate the
observed acceleration for a range of galaxy sizes (Lelli, McGaugh \& Schombert 2017).

To calculate the stellar mass of a galaxy ($M_*$) one requires 1) an accurate value
for the total stellar luminosity of a galaxy ($L_{tot}$) plus 2) a reliable
$\Upsilon_*$ at the same wavelength that the luminosity is determined.  While
photometry of galaxies still contains many inherent uncertainties (different aperture
sizes, foreground stars, nearby companion galaxies), the advent of areal detectors
and space imaging (where the sky brightness is substantially fainter) has removed
most of the limitations to assigning an accurate total luminosity to galaxies (i.e.,
an well defined curve-of-growth to the galaxy's luminosity profile).  Detailed
surface photometry also allows for a determination of the stellar luminosity per
parsec$^2$ and allows pixel-by-pixel evaluation of the underlying stellar population
(see Lee \etal 2018).  Thus, the uncertainties in stellar luminosity can be estimated
by various methods, but evaluating the reliability of an $\Upsilon_*$ is more
difficult due to the convolved path of stellar population modeling that is
involved in deducing the appropriate $\Upsilon_*$ at the wavelength of interest.

Two advances in recent years has dramatically improved our ability to deduce stellar
mass from photometry.  The first is the increased sophistication in the suite of
stellar isochrones used to produce stellar population models that allows inspection
of the effects of exotic components, such as horizontal branch and blue straggler
stars, as well as more detailed understanding of the effects of dust and the IMF.  In
addition, there is recent awareness that detailed SED fitting is not critical to
deducing $\Upsilon_*$ from population models, so that broadband photometry is
adequate (Gallazzi \& Bell 2009) as long as sufficient wavelength coverage is
obtained.  With this technique, one determines $\Upsilon_*$ through various
mass-to-light versus color relations (MLCR) as well as from detailed spectral
indices, 
a procedure that is laced with complications (see McGaugh \& Schombert 2014).

These improvements are well timed with the second advance on the observational side,
increasing numbers of nearby galaxies with detailed color-magnitude diagrams (CMD's)
of the resolved stellar populations plus improved photometry in the UV and near-IR
(i.e., $GALEX$ and $Spitzer$).  CMD's in nearby galaxies allow for direct comparison
between stellar population models and the actual stellar content.  The new
observations in the UV serve to constrain recent star formation rates, while new
near-IR observations provide more reliable luminosities to deduce $\Upsilon_*$ (as
$\Upsilon_*$ variations decrease significantly in the near-IR where old stars
dominate the baryonic mass, Rix \& Zaritsky 1995; Norris \etal 2016).

For star-forming galaxies, a valid $\Upsilon_*$ estimate requires two components, 1)
a reasonably constrained star formation history (knowledge of the change in the star
formation rate and chemical enrichment with time) and 2) reliable stellar population
models as a function of age and metallicity (to deduce how luminosities, as a
function of color, are converted into mass).  While it is straight forward to test
stellar population models against star clusters (Bruzual 2010) or quiescent galaxies,
like ellipticals (Schombert 2016), it is much more problematic to attempt to extract
colors and $\Upsilon_*$ from star-forming galaxies such as spirals and dwarf
irregulars (Bell \& de Jong 2000) due to the competing drivers of star formation and
chemical enrichment.  

There are two core observables for star-forming galaxies that can constrain their
complex histories and assist in deriving a stable $\Upsilon_*$'s; 1) the galaxy mass
versus SFR (star formation rate) diagram (the so-called main sequence for
star-forming galaxies, MSg, Noeske \etal 2007; Speagle \etal 2014) and 2) the locus
of galaxy's colors, particularly optical versus IR colors.  The MSg allows a crude
description of the star formation history (SFH); for most high-mass galaxies lie near
the region of gas exhaustion (meaning their past SFR must have been much higher than
the present), while low-mass galaxies lie near the line of constant SFR,
meaning their past SFR's must be near the current one (see Figure \ref{main_seq}).
This highly constrains the possible paths of past star formation as any past star
formation (SF) must be near the current SFR (although one can entertain many
scenarios, such as late formation epochs, see \S4).

In addition to the MSg, the suite of galaxy colors has also dramatically increased,
not only in number but in the range of wavelength's sampled.  Our focus, for this
study, is on the SPARC (Spitzer Photometry and Accurate Rotation Curves, Lelli,
McGaugh \& Schombert 2016) and S$^4$G (Sheth \etal 2010) datasets, as they have the
fullest coverage at 3.6$\mu$m, the $Spitzer$ channels, which are of highest interest
in deducing $\Upsilon_*$.  For a subset of several hundred galaxies in these two
samples, there is photometry from $GALEX$ $FUV$ to $Spitzer$ 4.5$\mu$m, including
optical ($SDSS$, RC3) and near-IR ($2MASS$, $WISE$).  The colors presented herein are
total colors, i.e., deduced from total luminosities in the various bandpasses.

The goal of this project is to combine new information from these color/SFR
relations, plus refined stellar population models, to obtain new color-$\Upsilon_*$
relations over a broad range of colors and stellar masses.  Our emphasis is on the
near-IR colors due their importance to the SPARC project (Lelli \etal 2017), however,
the models are applicable to a range of galaxy colors and are flexible to a range of
input parameters, such as different star formation histories and paths of chemical
evolution.

\section{Main Sequence Relationship for Star-forming Galaxies}

The main sequence for star-forming galaxies (where the designation of star-forming
includes, basically, all Hubble types later than Sa) is a somewhat surprising
relationship for galaxies since color and SFR vary widely with morphology suggesting
complicated star formation histories.  Considerable recent work has been motivated by
more accurate stellar mass determinations (using near-IR luminosities) and has
focussed on the star-forming main sequence through a direct comparison of the total
stellar mass of a galaxy versus its current SFR (Noeske \etal 2007; Salim
\etal 2007; Daddi \etal 2007; Peng \etal 2010; Wuyts \etal 2011; Cook \etal 2014;
Speagle \etal 2014; Jaskot \etal 2015; Cano-Diaz \etal 2016; Kurczynski \etal 2016).
The two parameters are clearly indirectly connected as the total stellar mass of a
galaxy reflects the integrated SFR over the complete SFH of the galaxy (or reflects
the SFH of the merger progenitors).  However, the current SFR only reflects the last
stage of an unknown, and possibly very complex, star formation history.  If star
formation has been a uniform process, then the current SFR presumingly scales with
the mean SFR of the galaxy and, thus, its integrated value becomes the total stellar
mass.  
A correlation between the current SFR value and total stellar mass, and its
narrowness, implies continuous evolution (Noeske \etal 2007) with roughly uniform,
ongoing SF during this time.  While the SFH's may have a range of shapes (e.g., on
and off bursts), the continuous consumption of gas implies that the evolution of SFR
with time must also be relativity similar across the various galaxy types (see also
Abramson \etal 2016).

We note that using the MSg to deduce $\Upsilon_*$ is somewhat circular, as total
stellar mass is deduced from total stellar luminosity, which we will then use to
constrain the value of $\Upsilon_*$ from stellar population models.  However, the
stellar mass axis is linear to changes in $\Upsilon_*$ and small changes to the total
stellar mass alters the integral value of the SFH.  Most of the observational
error is in the SFR values, and changes in the final SFR on the shape of the SFH has
a larger impact to the total stellar mass.

The term "main sequence" for galaxies is a poor analogue with the main sequence of
stars, which is driven by some basic nuclear physics, and the relationship between
stellar mass and current star formation (SF) has many competing processes to enhance
or suppress SF leading to complex histories behind the observed final outcome.
However, there are some interesting similarities.  
For example, the MSg relationship
is well defined with relatively low scatter on the low-mass end (Cook \etal 2014),
defined by the slowly evolving galaxies.  Considering the high gas fractions, typical
for low-mass LSB galaxies, this is surprising as one could imagine discontinuous
sharp bursts of SF (although not in line with their low stellar density appearances).
As one goes to higher stellar masses (lower gas fractions) the relationship displays
a "turn-off" at $M_* \approx 10^{10} M_{\sun}$ suggesting a point where gas depletion
occurs on timescales less than the age of the Universe ("weary giants", see McGaugh,
Schombert \& Lelli 2017).  The low-mass end contains the "thriving dwarfs" with their
plentiful gas supplies.

The main sequence relation plays a more important role when followed over redshift as
it then represents the SFR as a function of time per mass bin.  This has been
successfully applied by Speagle \etal (2014), who uses the change in the zeropoint of
the MSg to deduce an average SFH for galaxies (see their Figure 9).  That study,
notably, finds the slope of the MSg to vary only slightly with cosmic epoch and
defines the canonical fit for the current epoch ($z=0$) by extrapolation to the
current epoch.  
However, the fitted slope of the relation on the high-mass end varies
from rather flat at low redshift ($z=0.2$, Speagle \etal 2014) to rather steep at high
redshifts ($z=0.9$, Kurczynski \etal 2016).  We will use the shape of the
extrapolated $z=0$ SFH as the baseline shape of SFH for our analysis in \S4 (i.e., an
initial burst with a slowly declining SFR to the present epoch).
We make one small adjustment to the Speagle \etal SFH prescription in that we set the
initial epoch of star formation at 1 Gyr after the BB, rather than 4.  We find the
linear extrapolation from Speagle \etal Figure 9 to be in good agreement with our own
SPARC H$\alpha$ dataset (Lelli \etal 2016).  Nearly all the high-mass spirals must
follow some variation of this basic scenario in order to explain their positions on
the main sequence.

\begin{figure}
\centering
\includegraphics[width=\columnwidth,scale=0.80,angle=0]{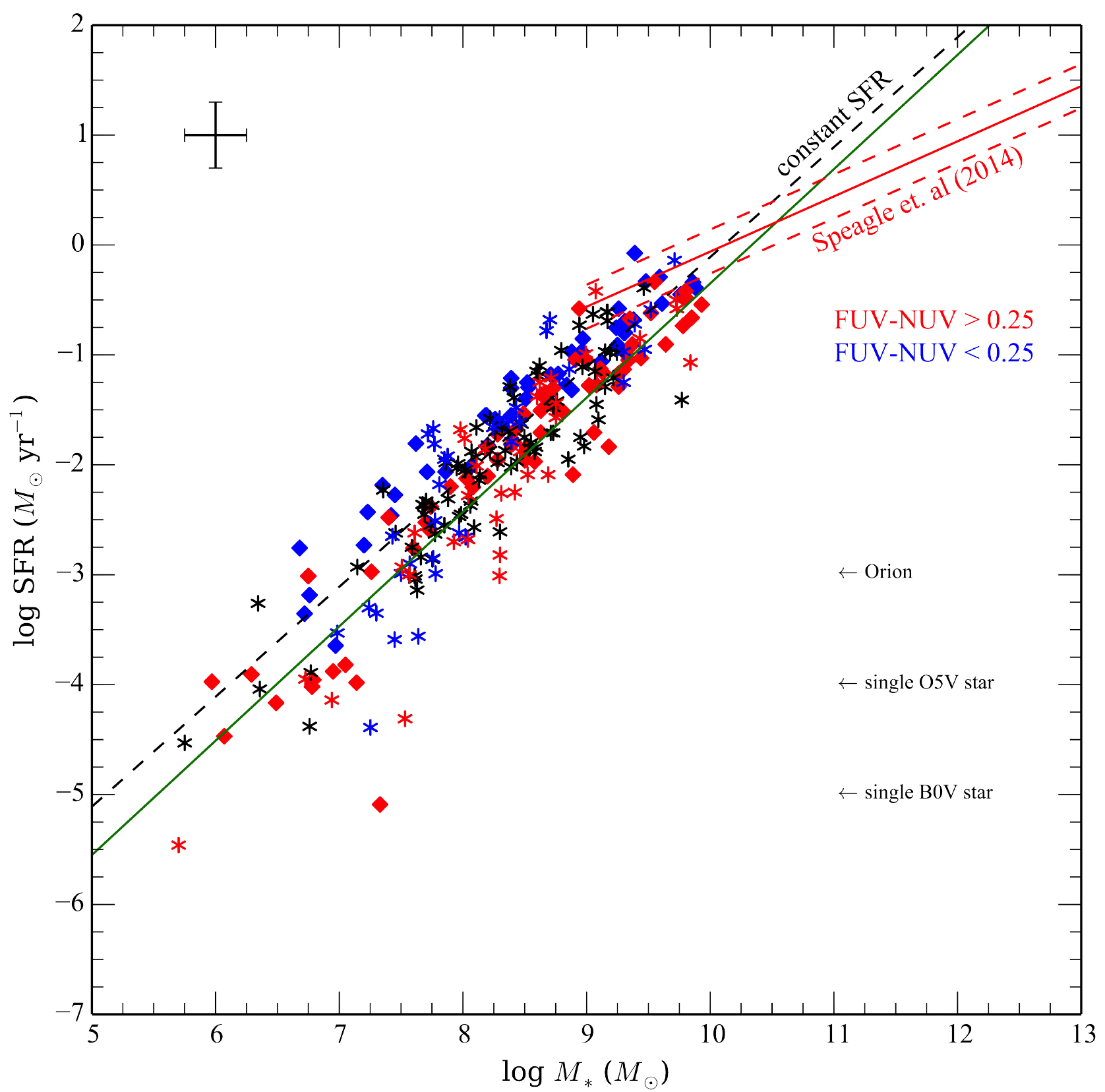}
\caption{\small The main sequence for high and low-mass star-forming galaxies.  The
datasets from Cook \etal (solid symbols) and LSB+SPARC (starred symbols) are shown, color
coded by $FUV-NUV$ color.  The green line is a fit to the LSB+SPARC sample (McGaugh,
Schombert \& Lelli 2017).  The dashed line is the line of constant star formation for
a 13 Gyr Universe.  There is a clear trend for blue $FUV-NUV$ colors to lie above the
constant SFR line (rising SFR in the last 100 Myrs) versus red $FUV-NUV$ colors below
the line (declining SFH).  The $z=0$ relationship from Speagle \etal is shown for the
high-mass spirals, along with 3$\sigma$ boundaries.  Also shown are the values for
SFR that correspond to an Orion-sized complex, a single O star and a single B star.
SFR estimates below $-$4.5 are highly inaccurate.  A representative error is shown in
the upper left, errors in SFR and stellar mass are from McGaugh, Schombert \& Lelli (2017).
}
\label{main_seq}
\end{figure}

However, the Speagle \etal (and most MSg studies) focus, primarily, on the high-mass
end, rarely fitting below $10^9$ $M_{\sun}$ (due to luminosity limitations of high
redshift samples).  On the low-mass end, a much steeper slope is found near a value
of 1 (Cook \etal 2014; McGaugh, Schombert \& Lelli 2017).  The data from Cook \etal
(2014) and the LSB+SPARC (a combination of our $Spitzer$ studies of low surface
brightness dwarfs plus the SPARC galaxies) dataset are shown in Figure \ref{main_seq}
along with the fits from McGaugh, Schombert \& Lelli (2017) plus the $z=0$ fit from
Speagle \etal (their equation 28).  A similar fit for low-mass disks was found by
Medling \etal (2018), confirming the downward turn of the MSg below log $M_*$ = 10.
The data is also flagged for $FUV-NUV$ color, where $GALEX$ data was available.  In
addition, the line of constant SFR for the age of the Universe (i.e., SFR $\approx
M_*/t_G$ for $t_G$ = 13 Gyrs) is also indicated for reference.

The advantage of the Cook \etal and LSB+SPARC datasets is that each determined the
current SFR by different methods.  The LSB+SPARC dataset used traditional H$\alpha$
observations to determine the current SFR using the canonical H$\alpha$-to-SFR
conversion of Kennicutt \& Evans (2012).  The Cook \etal sample used the $FUV$ flux
of a galaxy converted to the current SFR through the prescription of Murphy \etal
(2011).  For galaxies in common (15), they have a good one-to-one correspondence
between the deduced SFR, despite sampling slightly different timescales of SFR.  We
note that for the lowest mass galaxies, with log SFR $< -4$, that this corresponds to
a current SFR that is barely measurable (a single O star powered HII region).  At log
SFR = $-5$, the SFR is down to a single B star.  The $FUV$ flux is a better indicator
of SFR in the extremely low SFR realm since it covers a timescale of 100 Myrs and is
not dependent on the H$\alpha$ emission from short-lived (20 Myrs) O star complexes.
In either case, it is our opinion, that observed SFRs below $10^{-4.5}$ are extremely
inaccurate.

Two features are notable.  First, some downward extension below $M_* < 10^{10}$
M$_{\sun}$ is expected as the line of constant SFR intersects the Speagle \etal
relationship at that stellar mass.  But the Speagle \etal sequence for the weary
giants clearly deviates at $10^{10}$ $M_{\sun}$ from the slope defined by the Cook
and LSB+SPARC samples.  This is the region proposed by Peng \etal (2010) as the
transition from merger quenching versus mass quenching (a term to signify the many
feedback mechanisms that regulate/halt star formation).  Environmental effects also
begin to have a significant effect above this mass range (Speagle \etal 2014) plus
gas exhaustion, or strangulation, begins to flatten the main sequence above $10^{10}
M_{\sun}$ (Peng \etal 2015; McGaugh, Schombert \& Lelli 2017).

Second, the low-mass galaxies can be divided by their $FUV-NUV$ color, with the bluer
UV color galaxies having higher current SFR than the redder galaxies.  This divide
also occurs across the constant SFR line, signaling that galaxies with blue $FUV-NUV$
have rising SFR's (at least in the last 100 Myrs, the timescale measured by the $FUV$
flux) and red galaxies have declining SFR's in the last 100 Myrs.  Again, episodic
star formation can produce any total stellar mass and a broad range in current SFR.
The fact that low-mass dwarfs display a coherence in the last phase of SFR with their
position on the main sequence suggests a smooth, uniform SFH.

\section{Star-Forming Galaxy Colors}

The history of multi-color photometry of galaxies is lengthy (see Schombert 2018)
with the ultimate goal of using colors to untangle the underlying stellar populations
in galaxies.  While this has been successful for ellipticals, due to their simpler
SFH's and stellar populations colors dependent primarily on metallicity (Schombert
2016), star-forming galaxies present a more complicated interpretation of their
colors as age (i.e., SF) plays a dominant role.  Rather than attempting to deduce the
exact metallicity and age of the stellar populations in star-forming main sequence
galaxies (such as Bell \& de Jong 2000), a more promising path is to use their colors
to constrain the possible scenarios of star formation that produce their location on
the observed main sequence.  This allows an interpretation of $\Upsilon_*$, deduced
from stellar population models, as a function solely of galaxy color, with some
understanding of the scatter in $\Upsilon_*$ introduced by star formation assumptions
and folding variables, such as age and metallicity, into the integrated colors.

The relevant colors for star-forming galaxies, that impact on deducing their SFH's,
can be divided into UV (short of 3500\AA), optical (from 3500 to 5500\AA) and near-IR
(beyond 3 microns) colors.  There are numerous sources for optical colors ranging
from recent SDSS studies (Smolcic \etal 2006) to the RC3 (de Vaucouleurs \etal 1991).
The UV is dominated by results from $GALEX$ ($FUV$ and $NUV$, Morrissey \etal 2007).
The near-IR is sampled by $2MASS$ (Jarrett \etal 2000), $WISE$ (Wright \etal 2010)
and $Spitzer$ (Schombert \& McGaugh 2014).  For the purposes of exploring the
behavior of $\Upsilon_*$, $Spitzer$ 3.6$\mu$m data is preferred as it is farthest to
the red without encountering contamination by PAH emission.  

With the focus on $Spitzer$ 3.6$\mu$m photometry, we have also collected photometry
from the S$^4$G survey (Sheth \etal 2010) and combined this sample with our own
LSB+SPARC survey (Lelli, McGaugh \& Schombert 2016) and $FUV$ photometry from Cook
\etal (2014).  $Spitzer$ photometry was reevaluated for all three samples using
direct surface photometry of the images in the $Spitzer$ archive (see Schombert \&
McGaugh 2014 for description of the reduction pipeline).  After culling for
photometric accuracy (all galaxies had to have $Spitzer$ magnitude errors less than
0.3), 301 galaxies were extracted from the S$^4$G sample, 120 were extracted from
Cook \etal and 160 were extracted from the LSB+SPARC dataset.  Optical colors were
extracted from NED using a variety of sources plus our own optical photometry of LSB
galaxies in the SPARC sample (Pildis, Schombert \& Eder 1997).  The optical to
near-IR colors were made by comparing NED aperture magnitudes to the full curve of
growth in the $Spitzer$ photometry (using the largest aperture in NED).  A comparison
to SDSS DR14 images was made for the SPARC dataset (see Schombert 2016) to confirm
the NED optical values.  All magnitudes were corrected for Galactic extinction.

\begin{figure*}
\centering
\includegraphics[scale=0.99,angle=0]{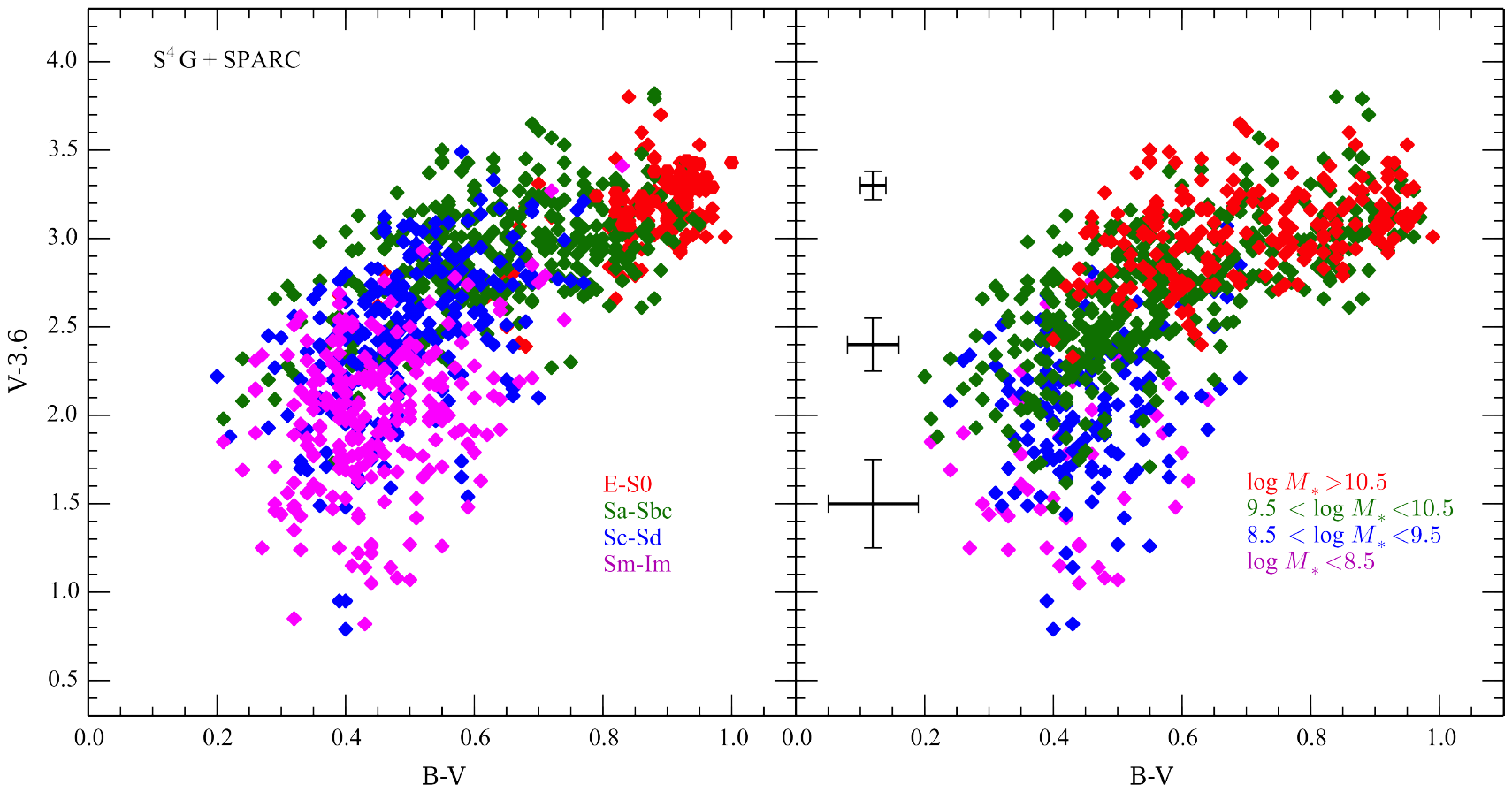}
\caption{\small The color locus of star-forming galaxies in optical $B-V$ versus the
near-IR $V-3.6$.  The left panel divides the S$^4$G and SPARC samples by
morphological type.  The right panel divides the samples by stellar mass, deduced
from 3.6$\mu$m luminosity.  All morphological types have a range of $B-V$ colors, but
spiral differentiate from dwarfs at a $V-3.6$ color of 2.5.  Stellar mass is
independent of optical color, but is better distinguished by $V-3.6$ color.
}
\label{two_color_side}
\end{figure*}

These three samples are displayed in Figure \ref{two_color_side}, corrected for
internal extinction (following the standard RC3 correction based on galaxy type and
inclination) and divided into four morphological classes; ellipticals/S0's,
early-type spirals (Sa to Sbc), late-type spirals (Sc to Sd) and late-type dwarfs
(Sm, Im, dI, etc.).  The trend for early-type galaxies to have redder colors is
obvious, although there is a great deal of mixing of color by morphological type.
This is due, primarily, to the fact that these are integrated total colors and, thus,
the blending of bulge and disk colors is unconstrained.  Morphological classification
by color is inaccurate, but we note there are very few late-type galaxies with
$V-3.6$ colors redder than 2.5 and few early-type spirals with $V-3.6$ colors bluer
than this line of demarcation.

The accuracy of the photometry varies based on the original source material (see a
similar plot with error bars in Schombert \& McGaugh 2014, Figure 6).  In general,
the error bars increase to bluer $V-3.6$ galaxies as these galaxies are lower in
surface brightness as a class (representative errorbars are displayed in Figure
\ref{two_color_side}).  The locus of color defined in Figure \ref{two_color_side} is
larger than the scatter in the observables and reflects a range of age and
metallicity paths for the underlying stellar populations (Bell \& de Jong 2000).  We
will use a density map of Figure \ref{two_color_side} for comparison to stellar
population models in \S6.

The right panel in Figure \ref{two_color_side} displays the same colors coded by
stellar mass (using a value of 0.5 to convert 3.6$\mu$m luminosities into mass, Lelli
\etal 2017).  The trend for higher masses to be redder follows the well known
color-mass and color-magnitude relationships (Tremonti \etal 2004).  The trend in
color, at first glance, is surprisingly coherent considering that the MSg finds the
highest mass star-forming galaxies to have the highest current SFR's and,
presumingly, largest number of young stars.  In other words, one might expect that
the galaxies with the highest SFR would have the bluest $B-V$ colors.  But this naive
interpretation ignores the fact that the past SFR must have been much higher in these
galaxies and, thus, in high-mass galaxies a majority of the stellar population is
composed of old stars with redder colors pushing $B-V$ to redder values (see Figure
\ref{fraction}).  This can also be seen by the fact that mass is more correlated with
$V-3.6$ color than $B-V$, optical colors having a larger range than near-IR colors
due to diversity in the SFH and variable internal extinction.

\section{Star Formation History}

Deducing the star formation history of a galaxy is a convoluted process that attempts
to extract the ages of the stars, by number, that make up its stellar population.  In
practice, this involves extracting the SFR as a function of time and applying a
standard initial mass function (IMF) to derive the total luminosity (i.e. stellar
mass) for the present epoch.  This is the approach used, successfully, by Speagle
\etal (2014) by following the main sequence as a function of redshift, effectively
measuring SFR as a function of lookback time then piecing together the SFH of a range
of galaxies by total mass (see also Leitner 2012).

\begin{figure}
\centering
\includegraphics[scale=0.50,angle=0]{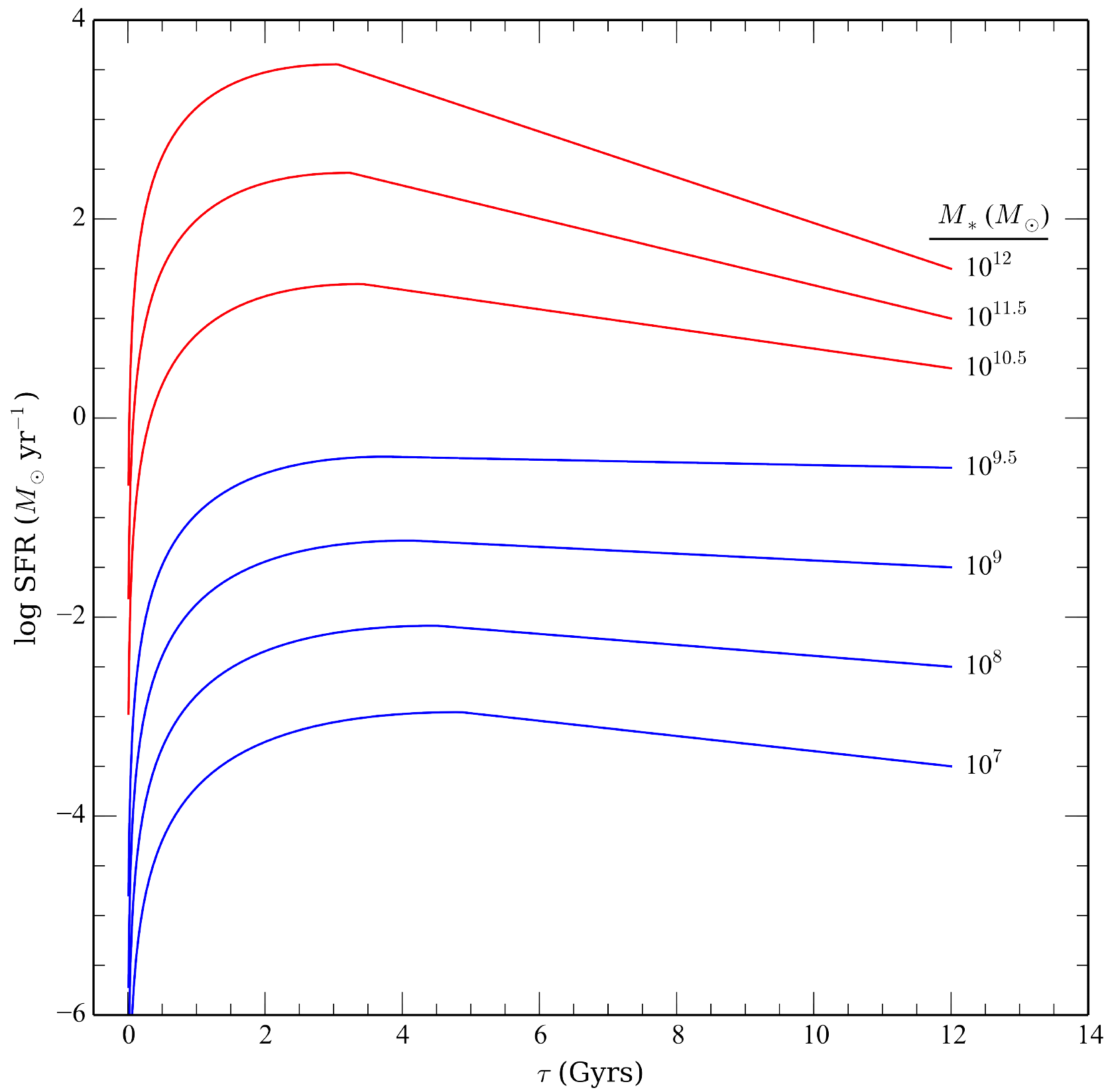}
\caption{\small Baseline star formation histories as a function of final SFR.  The
red curves are for high-mass ($M_* > 10^{10} M_{\sun}$) galaxies that follow Speagle
\etal $z=0$ main sequence.  The blue curves are adjusted to match the low-mass main
sequence found by McGaugh, Schombert \& Lelli (2017).  The shallower slope of the
Speagle \etal main sequence results in a sharper decline in SFR versus the low-mass
systems SFH.  These SFH's form the baseline for comparison in \S6, variations on
starting epochs, width of the initial burst and rapid changes as also considered.
}
\label{speagle_sfh}
\end{figure}

We can adopt Speagle's general SFH shape for star-forming galaxies greater than
$10^{10}$ $M_{\sun}$ (see their Figure 9).  However, the main sequence takes on a
different slope for lower mass galaxies suggesting a different form to the SFH for
these lower mass systems.  Our procedure starts with this general shape and the SFR
at the current epoch is the input.  This determines the total stellar mass through
the MSg.  The peak SFR is then normalized such that the integrated stellar mass from
this SFH matches the total stellar mass given by the MSg.  Figure \ref{speagle_sfh}
presents the Speagle \etal SFH, displayed as log SFR with respect to time from galaxy
formation ($\tau$).  The red curves follow the Speagle \etal shape, normalized such
that the final stellar mass agrees with their $z=0$ MSg.  Given the shallow slope of
the Speagle \etal main sequence, this naturally results in a shaper decline in star
formation for higher mass spirals.  

The SFH's for the lower mass end of the main sequence are shown as blue curves in
Figure \ref{speagle_sfh}.  In order to match the main sequence for low-mass galaxies
found by McGaugh, Schombert \& Lelli (2017), the strength of the initial burst must
be lowered with decreasing mass to a larger degree than the high-mass spirals.
However, the steeper slope of the MSg, near the line of constant SF, restricts the
strength of the initial burst in order to maintain nearly constant SF to the current
epoch (needed to match the final total stellar mass to the current SFR).  The
simplest solution is to extend the duration of the weak initial burst from 4 to 5
Gyrs from galaxies with 10$^9$ to 10$^7$ $M_{\sun}$.  There is some physical choice
for this alteration of the SFH with lower mass as it is well known that star
formation goes as the density of the gas (i.e., Schmidt's law).  Lower density
galaxies naturally have decreased SFR's; however, higher gas fractions
allows for longer durations of SF.  In any case, this is simply a numeral choice in
order to have the simplest parameters that integrate to the correct total stellar
mass given by the Speagle \etal SFH shape, plus end up on the correct position of the
main sequence for a given current SFR.  This prescription, unsurprisingly, results in
a nearly constant, SFR for low-mass galaxies, but also allows for slightly rising or
declining SFH (as indicated by the range in $FUV-NUV$ colors) and will be discussed
in \S6.  

While the general shape of the Speagle \etal SFH is applied to each mass bin to
produce final stellar masses in agreement with the observed main sequence, these
SFH's predict very different integrated properties.  This was first explored by
Leitner (2012) who uses the Main Sequence Integration technique, combined with high
redshift SFR information, to deduce similar SFH's to Speagle \etal (see his Figure 3)
and the mass fraction growth with lookback time.  For massive systems, Leitner finds
a majority of the stellar mass is in place within 2 to 3 Gyrs after initial SF.
However, to maintain the shallow slope of the high-mass MSg, his SFH have later
initial SF epochs with decreasing mass.  This also has the advantage of slowing the
chemical enrichment of low-mass systems, and produce bluer colors as the oldest stars
are only 6 Gyrs in age.  The steeper slope on the low-mass end of the MSg lowers the
current SFR below the extrapolated SFH used by Leitner and forces a longer duration of SF
in order to produce sufficient stellar mass.

\begin{figure}
\centering
\includegraphics[scale=0.50,angle=0]{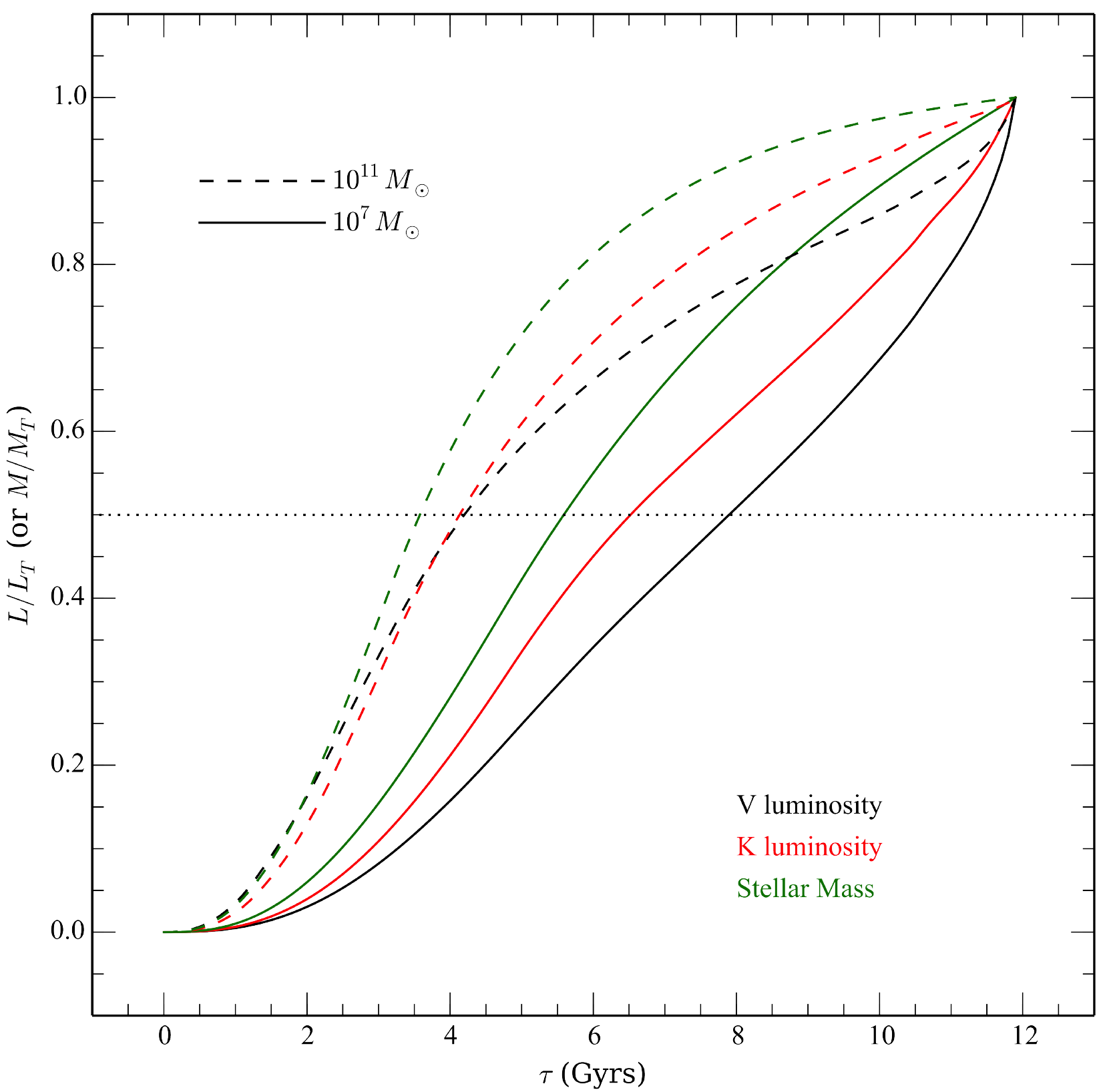}
\caption{\small For the baseline SFH in Figure \ref{speagle_sfh}, the resulting
growth in $V$ and $K$ luminosity plus stellar mass.  High mass spirals achieve a 50\%
line at much earlier epoch's than low-mass dwarfs, resulting in redder colors.  Slow
mass build-up also reflects into slow chemical enrichment for dwarfs, plus lower
final [Fe/H] values.
}
\label{fraction}
\end{figure}

We can compare the growth in stellar mass given by the fiducial SFH's in Figure
\ref{speagle_sfh}.  For example, in Figure \ref{fraction}, we can see that the growth
in stellar mass and luminosity differs significantly with increasing galaxy mass and
the contribution from stars of differing ages also varies significantly with
increasing galaxy mass.  A low-mass dwarf galaxy achieves 1/2 its total mass by 5
Gyrs from the start of star formation, compared to a high spiral which only takes 3
Gyrs to achieve the same fraction.  This results in a higher fraction of older stars
for high-mass systems.  A slower mass growth rate also reflects into the rate of
chemical enrichment, the lower metallicity of LSB galaxies is driven, not only by
lower SFR, but by overall slow-mass growth.  Based on these results, a different
chemical enrichment model will be considered for each mass bin (see \S5.1).

Using a standard stellar population model (see \S5), the $V$ and $K$ luminosities of
a low-mass dwarf reach the midpoint at 7 to 8 Gyrs, whereas a high-mass spiral only
requires 4 Gyrs.  This explains some of the difference in colors for LSB galaxies
versus bright spirals.  Despite the higher current SFR rates, most of the stellar
mass is locked in stars that are older than 6 to 7 Gyrs, a red population.  While
most of the stars in an LSB galaxy are younger than 6 Gyrs, their blue colors due to
younger mean age, not recent SF.  As we will see in \S6, growth in optical, compared
to near-IR colors, is also weak in high-mass spirals, producing redder $V-K$ colors,
versus LSB galaxies which have consistent differences in $V$ and $K$ luminosities
resulting in much bluer $V-K$ and $V-3.6$ colors.  Thus, the combination of low
metallicity, older populations and a stronger contribution from the younger stars 
is the primary reason that LSB galaxies occupy the bluest colors at all
wavelengths, not due to particularly recent formation epochs (Pildis, Schombert \&
Eder 1997; Schombert \& McGaugh 2014).

Unlike the high-mass end (the land of weary giants, running out of HI), the lower
mass end of the main sequence has a more diverse range in recent SFR changes as indicated
by their $FUV-NUV$ colors.  The line of constant star formation divides the lower
mass end of the main sequence plane into two sections; 1) those galaxies where
the average past SFR must have been higher than the current SFR (to the right of the
constant SFR line) and 2) those galaxies where the average past SFR must have been
lower than the current SFR (to the left of the constant SFR line).  All galaxies
higher in mass than $10^{10}$ $M_{\sun}$ lie to the right of the constant SFR line
and, thus, have declining SFRs in agreement with the deduced SFH from Speagle \etal
(2014).  However, the low-mass sample divides neatly by $FUV-NUV$ around the constant
SFR line (shown as red and blue symbols in Figure \ref{main_seq}).  

As shown by Murphy \etal (2011), dust-corrected $FUV$ luminosity is a proxy for the
current SFR rate on timescales of a few 100 Myrs (in contrast, H$\alpha$ luminosity
measures a shorter timescale, around 20 Myrs).  The mean $FUV-NUV$ color for
star-forming galaxies is 0.25 with a standard deviation of 0.10.  However, dividing
the sample by $FUV-NUV = 0.25$ displays a surprising dichotomy in Figure
\ref{main_seq}.  There are 187 galaxies with $FUV-NUV$ colors.  Of the $GALEX$
sample, 87 have $FUV-NUV <$ 0.25, 100 are redder.  Of the blue sample, 57 are to the
right of the constant SFR line (67\%), versus 25 of the red sample (25\%).  Thus, a
majority of star-forming galaxies with blue $FUV-NUV$ colors lie to the left of the
constant SFR line indicating that those galaxies have rising SFR's over that last few
100 Myrs.  Likewise, star-forming galaxies with red $FUV-NUV$ colors have declining
SFR's over the same timescale and lie to the right of the constant SFR line.  No
correlation between $FUV-NUV$ and $B-V$ is found, indicating that changes in SFR
occur on timescales of 100 Myrs, but are stable over Gyr timescales.  This is in
agreement with the deduced SFR over the last Gyrs in the WFC3 CMDs for three LSB
galaxies (Schombert \& McGaugh 2015).

\begin{figure}
\centering
\includegraphics[scale=0.50,angle=0]{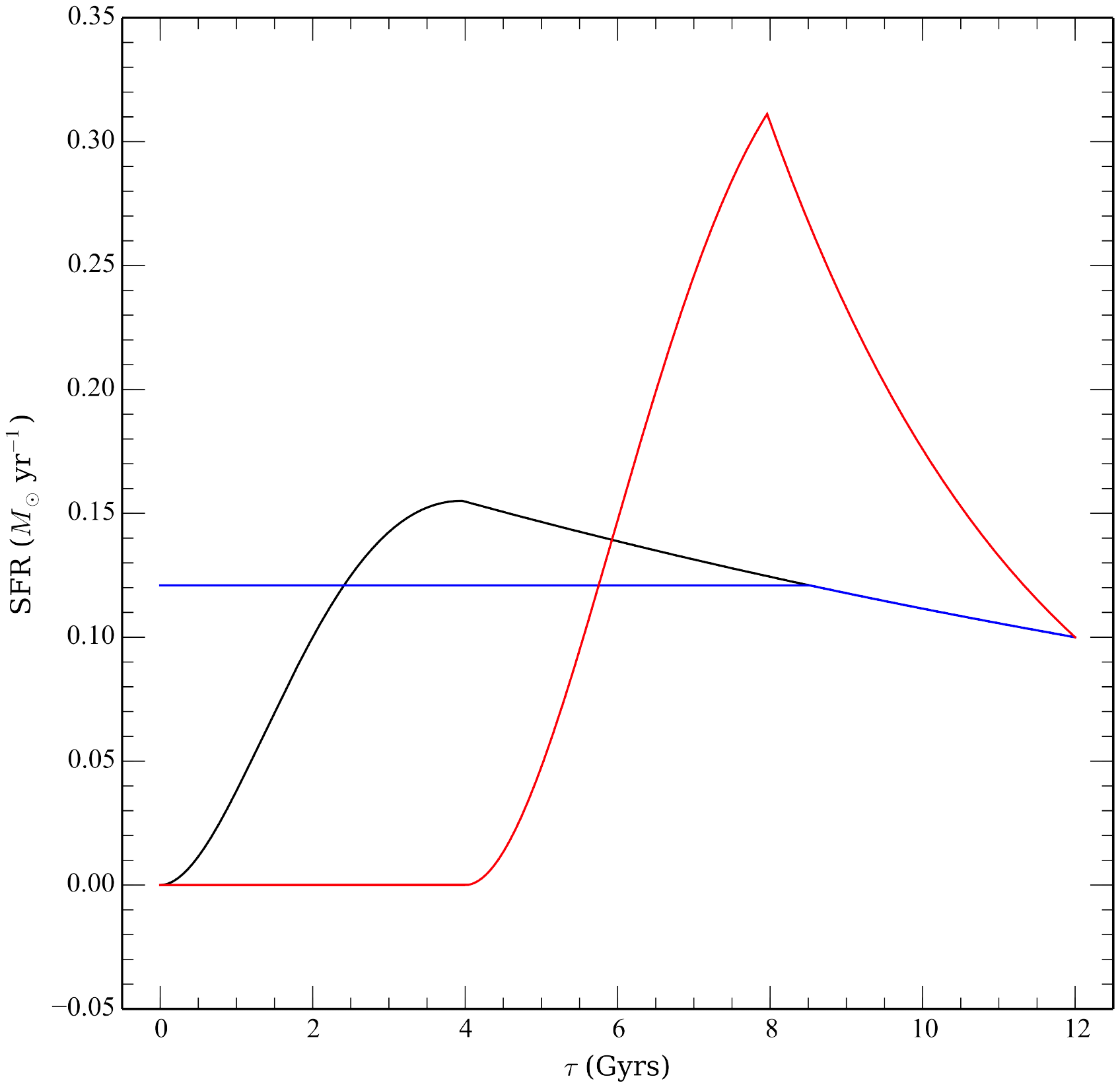}
\caption{\small Three possible SFH scenarios that reproduce the low-mass end of the
main sequence.  The black line displays the baseline SFH from Speagle \etal,
normalized such that the current SFR matches the total stellar mass produced by the
integrated SFH.  The red line is a late formation scenario, where initial star
formation is delayed by 4 Gyrs.  The initial burst must be increased to produce the
expected stellar mass.  The blue line is a scenario where the initial burst is
extended by several Gyrs, resulting in a lower peak burst.
}
\label{sfh_mix}
\end{figure}

The above dichotomy in $FUV-NUV$ colors indicates that the last stages in SFH
scenarios for low-mass dwarfs are sensitive to upward and downward changes in SFR.
The sample from LSB+SPARC straddle the constant SFR line, although a majority lie on
the right hand side indicating a declining SFR.  As the $FUV-NUV$ color correlates
with position with respect to the constant SFR line, we consider two avenues for the
SFH lower mass galaxies, one with a declining SFR of the last Gyr to its current
value, the second with a rising SFR over the last Gyr.  However, even with this
constraint, there are numerous potential SFH's that achieve the observed trend of
stellar mass with current SFR such that they 1) have the correct final mass for the
final SFR, 2) have recent SFR that are either greater than or less than the current
rate (depending on $FUV-NUV$ color) and 3) are declining or rising over the last Gyr
(again, defined by current $FUV-NUV$ color).  For example, Figure \ref{sfh_mix}
displays three possible scenarios that satisfy the above conditions for those
low-mass galaxies with declining SFR.  Each, scaled to the appropriate initial burst,
can reproduce the observed low-mass main sequence, but predict very different final
colors (see \S6).

In addition, we consider the effects of episodic star formation, particularly in
recent epochs (Noeske \etal 2007; Haywood \etal 2018).  In this scenario, we alter
the SFH's in Figure \ref{sfh_mix} with sharp changes in SFR over the last Gyr.
These small bursts must be fairly constrained otherwise there would not be a strong
correlation between $FUV-NUV$ and optical colors, such as $B-V$.  The scatter in the
$B-V$ versus $FUV-NUV$ diagram is 0.13 dex, where most of the uncertainty is due to
observational error on the colors.  A conservative estimate is that the SFR for
galaxies with $B-V < 0.7$ has not deviated more that 15\% in the last couple Gyrs
(the timescale sampled by $B-V$).   Similar estimates are obtained from comparison
with $U-B$ and $U-V$ colors.

The main sequence will not constrain any of the above scenarios.  However, each does
make specific predictions on the final integrated colors of galaxies.  Thus,
additional constraints can be obtained by using spectral energy distribution (SED)
models (Bruzual \& Charlot 2003; Conroy \& Gunn 2010) to predict present-day colors
from a given SFH history, then comparing this to the color locus of star-forming
galaxies binned by total stellar mass.  The input parameters (such as age and
metallicity distribution) are the critical unknowns in producing a reliable color
locus for star-forming galaxies.  Fortunately, new observations from $GALEX$ to
$Spitzer$ combined with HST CMD's of nearby LSB galaxies (Schombert \& McGaugh 2015)
serve to guide those inputs.

For the sake of numerical experiments in \S6, we divide the low-mass SFH scenarios
into five classes; 1) baseline (following the prescription deduced by Speagle \etal),
2) late (a formation epoch delayed by 4 to 5 Gyrs), 3) wide (an extended initial
burst of lesser intensity), 4) rising (a lower initial burst with a slightly rising
SFH) and 5) episodic (varying the late stages of SF by 20\%).  Each scenario is
mapped onto the main sequence as a boundary condition to the total level of star
formation (shown in Figure \ref{sfh_mix}) with some adjustment to consider rising or
declining star formation based on the division by $FUV-NUV$ color.  Each scenario can
then be mapped into a stellar population plus chemical enrichment model, as discussed
in the next section.

\section{Stellar Population Models}

A previous paper by the SPARC team (Schombert \& McGaugh 2014) derived the
$\Upsilon_*$ ratio of star-forming LSB galaxies based on experience with elliptical
colors and the SED models from Bruzual \& Charlot (2003, hereafter BC03).  In that
paper, empirical color relationships were used to extended the behavior of the BC03
models to farther wavelength's, particularly the $Spitzer$ 3.6$\mu$m filter and also
considered the effects of AGB, BHB and lower metallicity stars on the integrated
colors in a semi-empirically fashion.  While that technique was sufficient to derive
$\Upsilon_*$ for simple star formation histories, the more complicated paths outlined
in \S4 require a more sophisticated treatment.  Thus, we turn to the SED models
produced by Conroy \& Gunn (2010, hereafter CG10) which allow more flexibility in the
contribution of exotic populations (such as AGB and BHB stars) than the BC03 models,
and offer colors from $GALEX$ to $Spitzer$.  For standard stellar populations
assumptions (i.e., the same IMF and isochrones) both BC03 and CG10 agree on both
optical and near-IR colors, demonstrating consistence in their techniques.

Our procedure is similar to that outlined in Schombert \& McGaugh (2014), each model
is the sum of a number of simple stellar populations (SSPs) which consists of a unit
mass of stars of the same age and metallicity.  Each SSP is initialized with a
selected IMF and each stellar mass bin has been evolved to a set age using a suite of
stellar isochrones (see CG10 or BC03 for details of the stellar libraries).  At any
particular age, the sum of all the previous SSP's (we use timesteps of 0.01 Gyrs)
weighted by the luminosity of the stellar population is output either as a single
spectra, or convolved through a set of standard filters.  By integrating their entire
spectrum of each SSP we can achieve a goal, as noted by Roediger \& Courteau (2015),
that increasing wavelength coverage significantly reduces the uncertainty and
systematic errors in $\Upsilon_*$ and, thereby, stellar mass estimates.  The scripts
to perform these calculations are available at our website.

While the evolution of a stellar population using isochrones is a stable calculation,
there are numerous subtleties to the detailed calculations.  For example, the IMF
used to make the initial mass distribution can vary (e.g., a Salpeter 1955, Kroupa
2001 or Chabrier 2003 style).  Metallicity is straight forward, but variations of the
$\alpha$ element ratio (typically expressed as $\alpha$/Fe), driven by the time
sensitive ratio of Type Ia to Type II SN, effects the number of free electrons in a
stellar atmosphere, which in turns alters the temperature of the RGB.  The
contribution of blue horizontal stars (BHB) and blue stragglers (BSs) is a free
parameter as is the treatment of thermal pulsing asymptotic giant branch (TP-AGB)
stars.  The former being important for optical colors, the later dominating the
near-IR colors.

Globally, the relevant inputs are the star formation history plus a chemical
enrichment model.  These allow each individual SSP to be summed over the SFH with
each SSP using the [Fe/H] value at each epoch as given by the enrichment model.  The
check on the final result will be the models position on the main sequence (stellar
mass versus current SFR), reproducing the mass-metallicity relation (the stellar mass
versus average [Fe/H] value for the model), the internal metallicity distribution
(compared to the MW, e.g., the G dwarf problem) and the integrated model colors.  All
the details of gas infall, recycling and stellar remnants are contained in the
chemical enrichment models, so this will be discussed first.

\subsection{Chemical Enrichment Model}

The basic assumption behind a basic chemical evolution scenario is instantaneous
recycling of enriched material by mass loss or supernovae ejecta (Pagel 1997).  While
it does take a finite amount of time to process the ISM through the birth and death
of stars, fortunately, for the calculations, the recycling timescales are much
shorter than the timesteps used for our SFH models (Matteucci 2007).  The one
exception to this rule is the effect of the $\alpha$/Fe ratio which can require Gyrs
to build-up and will be discussed in \S5.2.

For the galaxies we attempt to model, we are guided by the age-metallicity scenarios
proposed by Prantzos (2009) based on an analysis of the stars in the Milky Way.  The
scenario has several common features with observations of MW stellar populations; 1)
a pre-enriched population with [Fe/H] $= -1.5$, 2) a rapid rise to approximately 80\%
of the final metallicity in 5 Gyrs, then 3) a slow rise to the final [Fe/H] of
the current epoch.  A small adjustment to the model is made to account for the lower
percentage of metal-poor stars than predicted by the models (i.e., the G-dwarf
problem, see Schombert \& Rakos 2009).

Initial comparison between the models and CMD diagrams of nearby galaxies suggests
that the Prantzos scenario for the MW over-estimates the rise in metallicity for
low-mass galaxies and under-estimates the enrichment rate for high-mass galaxies.  To
produce a more realistic model, we alter the form of the Prantzos scenario into three
types; slow, normal and fast.  The slow version is basically a linear increase in
[Fe/H] with age, simulating a LSB dwarf with very slow SFRs.  The fast version is a
slightly more rapid rise (reaching 80\% final metallicity in 2 Gyrs, rather than 5 to
simulate a strong initial burst, as expected for high-mass, high SFR spirals.  Given
that we know, due to the steep slope of the main sequence below $10^{10}$ $M_{\sun}$,
that low-mass, LSB galaxies have either slowly rising or slowly declining SFH's, a
slower chemical enrichment seems appropriate.  We adopt the normal Prantzos scheme
for MW sized spirals, and the faster scheme for high-mass spirals, again based on the
slope of the main sequence.

As an external check to the merit of our enrichment prescriptions, we examined the
internal metallicity distribution functions (MDF) of the final stellar populations
and compared these to observed MDF's in ellipticals, spirals and the Milky Way
(Haywood \etal 2018).  We found that the shape of the MDF's were similar to the MW
MDF (including the G dwarf deficiency) and, as demonstrated in Schombert \& Rakos
(2009), the importance of low metallicity stars decreases with lower mean
metallicities as the MDF's compress in metallicity range.  For [Fe/H] values below
$-$0.5 (approximately 10$^9$ $M_{\sun}$ on the main sequence) the shape of the MDF
was irrelevant to the colors of the integrated stellar population.

Lastly, we need to assign a final [Fe/H] to each enrichment model.  We are guided by
the various mass-metallicity relation studies using O/H values for star-forming
galaxies (see Tremonti \etal 2004, Zahid \etal 2011 and Brown \etal 2018) First, the
relationships from these studies can be extrapolated to stellar masses between $10^8$
and $10^{11}$.  We convert the deduced O/H values into [Fe/H] using the prescription
from McGaugh (1991).  Second, for stellar masses below $10^8$ $M_{\sun}$, we use the
[Fe/H] values which have been measured directly for three LSB dwarfs (see Schombert
\& McGaugh 2015).  These dwarfs, with stellar masses around $10^7$ $M_{\sun}$, have
[Fe/H] values between $-$1.0 and $-$0.6 with a mean value of $-$0.7 at stellar masses
of 10$^7$ $M_{\sun}$.  Thus, we assume a [Fe/H] value of $-$1.2 for our lowest mass
galaxies, rising to a value of $-$0.5 where O/H values are available around stellar
masses of $10^{8.5}$ $M_{\sun}$.

\subsection{$\alpha$/Fe Corrections}

The colors of stellar populations are determined by the distribution of stars, given
by stellar isochrones, in the HR diagram.  Decreases in metallicity drive both the
turnoff point and the position of the RGB to hotter temperatures, i.e. bluer colors,
due to changes in line blanketing and opacity.  While Fe is the main contributor of
electrons to produce color changes, all atoms heavier than He can contribute
electrons.  It is typically assumed that all the elements track Fe abundance, but it
is possible to have overabundances of various light nuclei (so-called $\alpha$
elements) under certain conditions.

The $\alpha$ elements, everything lighter than Fe, are primarily produced in massive
stars (Type II SN), while a higher contribution to Fe comes from Type Ia SN.  Since
SNII are short-lived ($\tau < 10$ Myrs) and SNIa detonate only after a Gyr (the
average time for the white dwarf to form), the ratio of $\alpha$/Fe measures these
different formation timescales.  At early epochs, for a stellar population undergoing
constant SF, the ratio of $\alpha$/Fe is high as determined by Type II supernovae.
Then, after a Gyr, the $\alpha$/Fe ratio begins to decrease due to the contribution of
products from SNIa explosions (c.f., McWilliam 1997, and references therein).

For example, in elliptical galaxies the ratio of $\alpha$/Fe is a factor of four
higher than metal-rich stars in the Milky Way due to the short timescales of initial,
very strong bursts star formation (Thomas \etal 2004).  As star formation is extended
in star-forming disks and irregulars, those galaxy types have lower $\alpha$/Fe
ratios as more recent star formation is enriched in Fe from Type Ia SN (Matteucci
2007).  For a constant star formation scenario, we can model the ratio of $\alpha$/Fe
as a function of [Fe/H] following data from the Milky Way (Milone, Sansom \&
Sanchez-Blazquez 2010) with [Fe/H] serving as a proxy for age.  In their data,
elliptical-like $\alpha$/Fe ratios are found up to $[Fe/H]=-1.0$, then drops quickly
to a solar value at solar metallicities.  We can apply the same behavior to our
models.

The effect of the $\alpha$/Fe ratio on colors was outlined in Cassisi \etal (2004).
With respect to colors, $B-V$ decreases (bluer) with increasing $\alpha/Fe$, for
example a metal-poor population ([Fe/H]=$-$1.3) had a $B-V$ shift of $-$0.03 for an
increase in $\alpha$/Fe by a factor of four.  A solar metallicity population is
shifted by $-$0.07 blueward for the same change in $\alpha$/Fe.  Similar shifts are
expected in $V-K$ with $\Delta(V-K)$ ranging from $-$0.06 for metal-poor populations
to $-$0.09 at solar metallicity.

To incorporate these corrections into our models, we assume that $\alpha$/Fe
decreases, in a linear fashion (as Fe increases from Type Ia SN events), from an
initial value of 0.4 to a solar value (0.0) over one Gyr of time starting two Gyrs
after initial star formation.  This mimics the behavior seen in ellipticals and,
thus, only the first three Gyrs of star formation have differing $\alpha$/Fe values
from solar.  For star-forming galaxies less that $10^9$ $M_{\sun}$, the assumed
chemical enrichment scenario is much slower than what we used in Schombert \& McGaugh
(2014).  This would serve to decrease the impact of an $\alpha$/Fe correction as most
of the stars with high $\alpha$/Fe ratios are still quite low in metallicity
(typically below [Fe/H] $=-1.0$).  For slowing rising or slowing declining SFR, less
than 25\% of the total stellar population requires an $\alpha$/Fe correction.  
Our initial experiments indicated that this correction is very small and we will only consider
this correction in the color error budget (see \S6).

\subsection{Dust}

For this study, we follow the phenomenological dust models used by Conroy \& Gunn
(2009).  Dust affects galaxy colors through three avenues; 1) dust specific to young
stars (remnants of the stellar nursery), 2) circumstellar dust associated with AGB
stars, and 3) general dust attenuation from the diffuse ISM, which reddens integrated
colors with decreasing strength to longer wavelengths.  Dust attenuation for young
stars arises due to the dust in the molecular clouds in which very young stars are
embedded.  The timescale for this attenuation is $10^7$ years (Charlot \& Fall 2000)
and only applies to the last timestep of our models.  In addition, at log SFR $< -2$,
the luminosity from young stars makes this correction negligible.  Circumstellar
dust around AGB stars can play a significant role in the colors of that short-lived
population.  It is a metallicity dependent factor, but can be tested for in our
models by comparison of near-IR colors of LMC/SMC young clusters (see \S6).

General dust attenuation, generated by the diffuse ISM, is best handled using a
phenomenological model that is not computationally expensive (Conroy, White \& Gunn
2010).  In this case, an SSP of a set age has an attenuation curve with a age
dependent shape and normalization.  As the grain properties of dust are dependent on
mean metallicity (Guiderdoni \& Rocca-Volmerange 1987), the effects of attenuation
also decreases with lower [Fe/H] for the stellar population (assumed to be one of the
reasons low metallicity LSB galaxies have never been found to have dust features nor
far-IR emission).  Also, unlike extinction due to dust in the Milky Way, dust in
external galaxies scatters blue light which, on average, reenters the line-of-sight
(Calzetti 2001).  
This works to minimize any corrections for dust, however, again, we
will note this effect in our error budget and find the corrections for the low-mass
end of the MSg to be negligible.

\subsection{BSs and BHB Populations}

Two exotic stellar populations can play an important role for optical colors, blue
horizontal branch (BHB) stars and blue straggler stars (BSs). Horizontal branch stars
are old, low-mass ($M < 1 M_{\sun}$) stars which have entered the helium core burning
phase of their lives.  They are bright ($M_V = -5$), of nearly constant luminosity
and range in color from red to blue (RHB and BHB stars).  BHB stars are of interest
to galaxy population models for they have similar characteristics to young stars in
color parameter space, although they are not a signature of recent star formation.

Blue stragglers stars (BSs) occupy a position in the HR diagram that is slightly
bluer and more luminous than the stellar populations main sequence turnoff point
(Sarajedini 2007).  Their extended main sequence lifetime appears to be due to binary
mass exchange, either by close contact binaries (McCrea 1964) or direct collision
(Bailyn 1995).  Their importance to stellar population synthesis is that they occupy
a region of the HR diagram that mimics star formation and low metallicity effects
(i.e., increased contribution to the blue portion of the integrated SED).

The effect BHB stars on population models is limited in time and metallicity.  BHB
stars are primarily found in metal-poor clusters ($[Fe/H] < -1.4$) and are not found
in any population younger than 5 Gyrs.  Very few BHB stars are found in the solar
neighborhood (Jimenez \etal 1998), presumingly a combination of young age and high
metallicity, so their contribution in field populations is unclear.  Considering
low-mass galaxies with a slow chemical evolution then, at the start of a constant
star formation scenario, 50\% of the stellar population has the metallicity and age
appropriate for a BHB phase.  Following the prescription of Conroy, Gunn \& White
(2009), this corresponds to a $f_{BHB}=0.5$ which translates into a $\Delta(B-V)$ of
$-$0.05 and a $\Delta(V-K)$ of $-$0.03 for a low metallicity SSP ($[Fe/H] < -1.0$).
For a SFH dominated by a strong initial burst, this factor would be slightly larger
(up to 60\% of the final stellar population).  For a rising SFH, this effect will be
smaller (down to 25\%).

With respect to the BSs population, a more general problem of binary star evolution
is outlined in Li \& Han (2008) which takes into account binary interactions such as
mass transfer, mass accretion, common-envelope evolution, collisions, supernova
kicks, angular momentum loss mechanism, and tidal interactions (Hurley, Tout \& Pols
2002).  The results from those simulations indicate that, while BSs's are difficult
to model and relatively time sensitive, they are similar to BHB stars in that they
only contribute after a well-formed turnoff point develops at 5 Gyrs.  If collisions
are important for their development, then they will be more rare in galaxy stellar
populations due to lower stellar densities compared to globular clusters.  They appear
to be numerous in the Milky Way field populations (Preston \& Sneden 2000), but are
an order of magnitude less luminous than BHB stars.

The simulations of Li \& Han (2008) display a maximum of $-$0.03 bluer colors in
$B-V$ and $-$0.10 bluer colors in $V-K$ for populations older than 5 Gyrs.  The
spread in metallicity is small for $B-V$, approximately 0.01 and non-existent for
$V-K$.  Again, following the prescription of Conroy, Gunn \& White (2009), the same
level of significance for BSs as for BHB stars results in only a $\Delta(B-V)$ of
$-$0.02 and a negligible effect on $V-K$.  This is mostly due to the lower luminosity
of BSs populations.

For older populations, it appears that BHB stars dominate in luminosity over BSs
stars simply based on the fact that BHB corrections to SSP and elliptical narrow band
colors are sufficient to reproduce the color-magnitude relation (Schombert \& Rakos
2009).  For the scenario of constant star formation, by the age where BHB stars
decrease in their contribution ($\tau < 5$ Gyrs), BSs stars would be beginning to
influence the bluest wavelengths. However, young stars quickly overwhelm the BSs
luminosities and our simulations indicate that the BSs contribution is negligible
when compare to other factors.

\subsection{TP-AGB Treatment}

An important component for the near-IR colors is the treatment of thermally-pulsating
AGB stars (TP-AGBs or just AGB).  These are stars in the very late stages of their
evolution powered by a helium burning shell which is highly unstable.  They are stars
with high initial masses ($M > 5 M_{\sun}$) and intermediate in age ($\tau > 10^8$
years).  While the BC03 codes (and their extension, see Bruzual 2009) include
AGBs as part of their evolutionary sequence, comparison with other codes (e.g.,
Maraston 2005, Gonzalez-Perez \etal 2014) finds discrepancies in the amount of
luminosity from this short lived population.

\begin{figure}
\centering
\includegraphics[scale=0.50,angle=0]{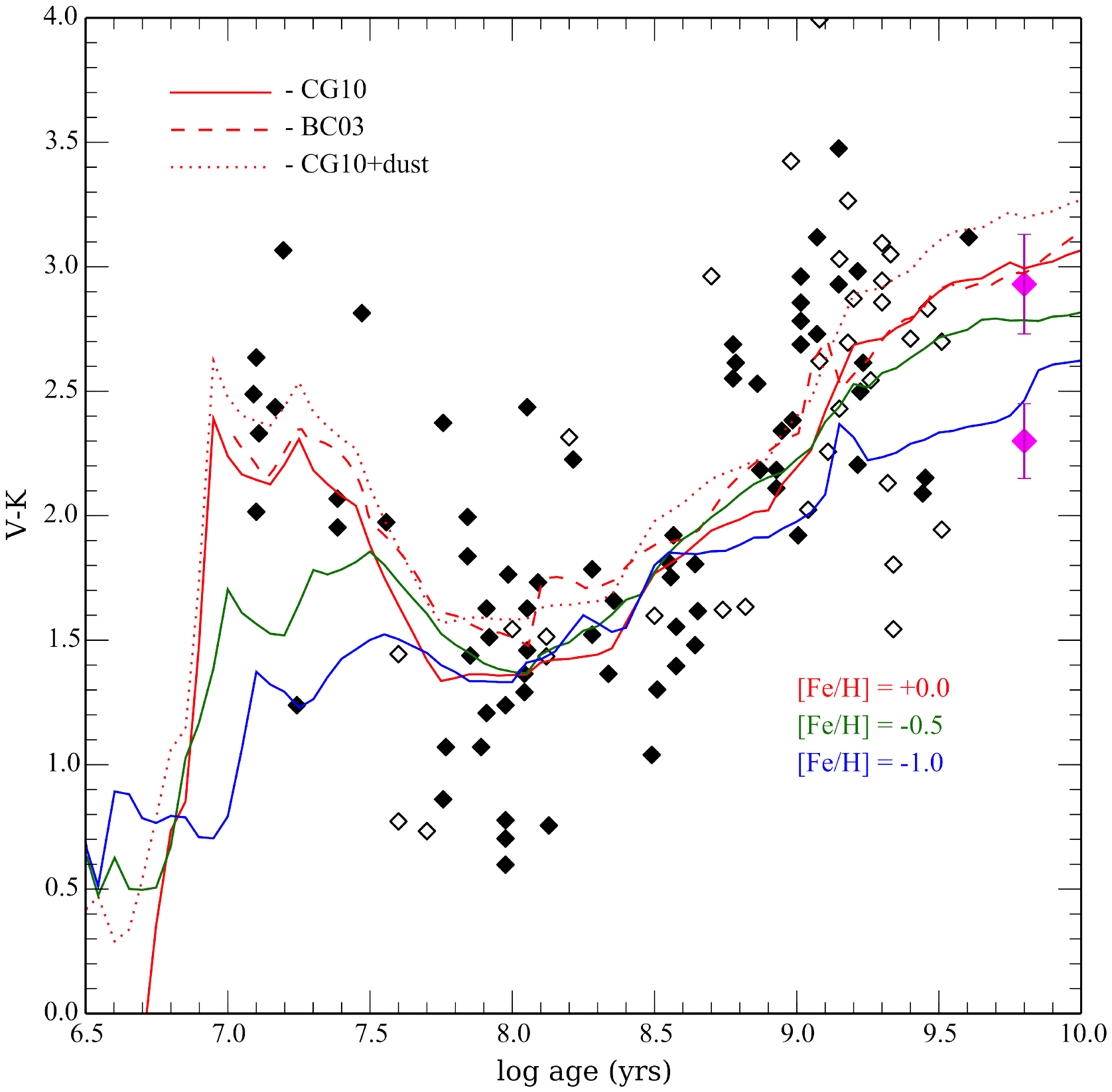}
\caption{\small $V-K$ colors of LMC/SMC star clusters from Kyeong \etal (2003, solid
symbols) and Pessev \etal (2006, open symbols).  Averages for [Fe/H] $< -1.0$ and $>
-1.0$ MW globulars from Cohen \etal (2007) are shown as magenta symbols.  The red
lines are solar metallicity SSP's from BC03 and CG10, with and without dust as
indicated.  The solar metallicity models are a good match to the young clusters, but
a poor match to the metal-poor older clusters.  Lower metallicity SSP's (green and
blue) reproduce the intermediate age clusters, but also fail to match the redder
colors of the oldest clusters.
}
\label{mc_clusters}
\end{figure}

The history of AGB treatment in SED codes is outlined in CG10.  To determine which
model best fits our AGB prescription, we compare the extension of BC03 and CG10 to
the near-IR colors of LMC and SMC star clusters (see Figure \ref{mc_clusters}).  As
metallicity typically decreases for older star clusters, it is problematic to compare
a single metallicity SSP track to the colors versus age in Figure \ref{mc_clusters}.
However, as can be seen in Figure \ref{mc_clusters}, a solar metallicity model
accurately captures the young cluster colors.  There is very little difference
between the solar BC03 and CG10 tracks, although adding a standard AGB dust model
(Villaume \etal 2015) compares more favorably with the redder young cluster colors.
The variation in metallicity is only important for very young and very old clusters.

As noted in Schombert (2016) and Schombert \& McGaugh (2014), all the models were
poor matches to the near-IR colors of populations older than a few Gyrs.  We will
still consider an empirical enhancement to model $V-K$ and $V-3.6$ colors for old,
metal-poor components.  However, comparison between metal-poor ([Fe/H] $< -1.0$) and
metal-rich ([Fe/H] $> -1.0$) MW globulars (magenta points in Figure
\ref{mc_clusters}) finds that the CG10 and BC03 models accurately predict their $V-K$
colors.  The discrepancy between model and observations occurs, primarily, for the
clusters with ages near $10^9$ yrs.

In a recent development, any corrections from our original paper (Schombert \&
McGaugh 2014) are now drawn into question.  Not due to changes in SED models, but
rather the recent HST observations on three, nearby LSB dwarfs (Schombert \& McGaugh
2015).  The F555W-F814W CMD's for LSB galaxies F415-3, F608-1 and F750-V1 display a
dramatic lack of AGB stars (only 5 to 10\% compared to 20\% for other metal-poor,
high SFR dwarfs).  A lack of AGB stars was doubly surprising as the metallicities of
these LSB dwarfs ([Fe/H $\approx$ $-0.8$) was in the realm where AGB stars have a
stronger contribution than for higher metallicity dwarfs.  It appears that the
extremely low past SFR for LSB galaxies (log SFR $< -3$) combined with the short
lifetimes for the AGB population under produces, by a significant fraction, the
expected numbers from simple SFH scenarios.

To account for this deficiency, we use the CG10 models ability to alter the
contribution from AGB stars, and allow for an underabundance, or overabundance, to be
considered combined with a metallicity threshold.  Thus, we consider two AGB
corrections, one where the AGB component is suppressed, as indicated for low-mass
dwarfs from WFC3 observations, and a second scenario where the near-IR luminosities
are slightly boosted between 0.1 and 2 Gyrs in concurrence with LMC/SMC cluster
observations (Ko, Lee \& Lim 2013).

\subsection{Empirical Calibration of Near-IR Colors}

\begin{figure}
\centering
\includegraphics[scale=0.50,angle=0]{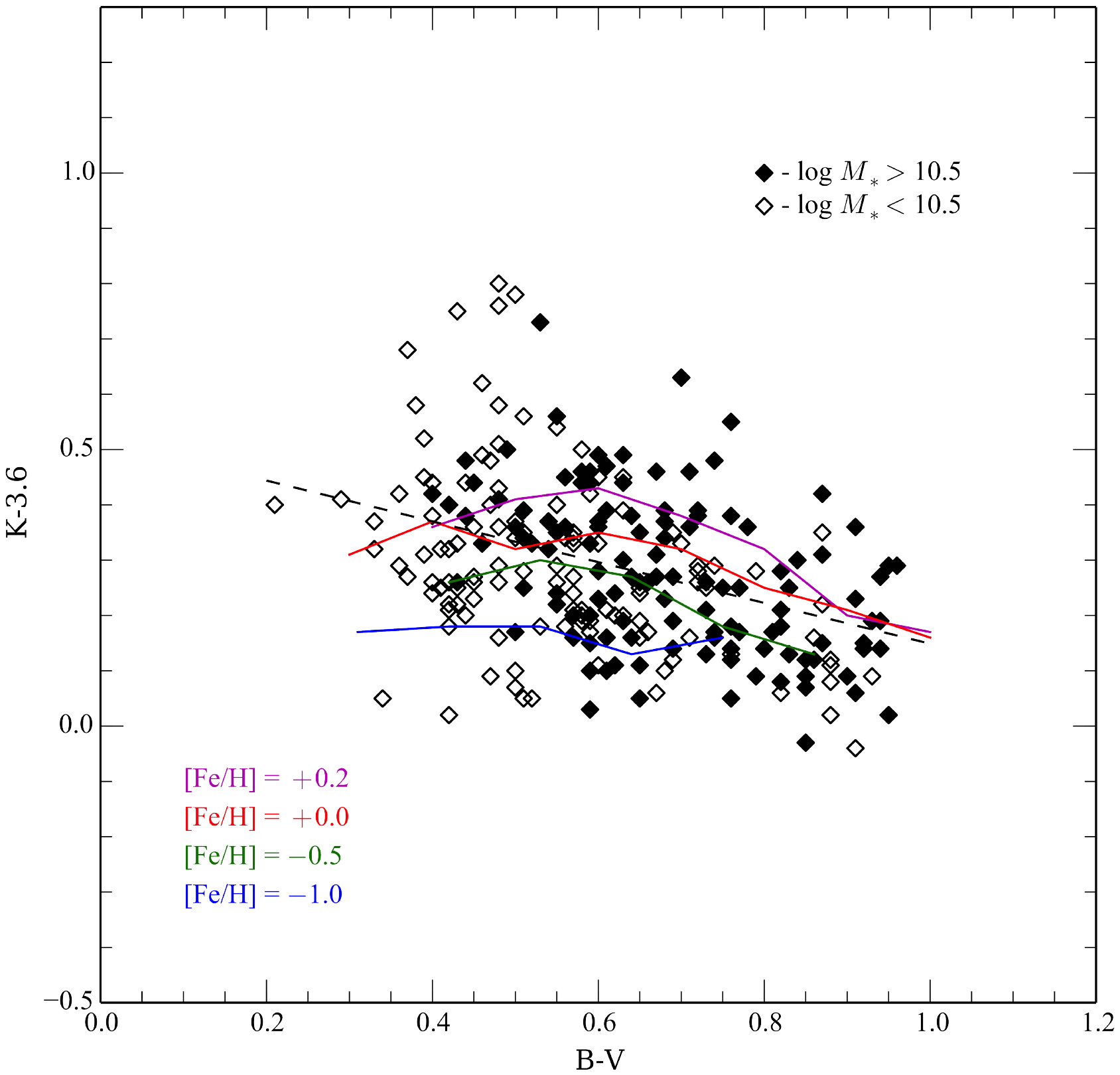}
\caption{\small The two color diagram for $B-V$ versus $K-3.6$ using the SPARC plus
S$^4$G samples.  A least-squares fit is shown and symbol type divides the sample into
high and low-mass systems (divided at $10^{10.5}$).  As found by Schombert \& McGaugh
(2014), the mean $K-3.6$ color is around 0.3 with a slight color term such that
optical bluer galaxies have slightly redder $K-3.6$ colors.  Also shown are four SSP
models from CG10 for [Fe/H] values of +0.2, 0.0, $-$0.5 and $-$1.0 (more metal-rich
have redder $K-3.6$ colors).  The SSP's range from $10^8$ to $10^{10}$ years in age.
}
\label{bv_k36}
\end{figure}

The SPARC dataset depends on $Spitzer$ 3.6$\mu$m observations to determine the
stellar mass which, when combined with the gas mass, becomes the total baryon mass of
a galaxy.  An accurate $\Upsilon_*$ requires a valid stellar population model at the
$Spitzer$ wavelengths, however, the BC03 models do not extend beyond $K$ (3.2$/mu$m).
To correct for this, Schombert \& McGaugh (2014) used an empirical $K-3.6$
calibration, based on $2MASS$, $WISE$ and $Spitzer$ observations, to convert BC03
$\Upsilon_*^K$'s into $\Upsilon_*^{3.6}$.  In the interim, improved $K$ observations
and additional $Spitzer$ observations have been obtained for the SPARC sample.  When
combined with existing datasets, such as S$^4$G (Sheth \etal 2010), this allows for a
detailed comparison with the CG10 models, which use an extended stellar library to
offer 3.6$\mu$m colors, and near-IR colros.  Figure \ref{bv_k36} displays the two
color diagram for $B-V$ versus $K-3.6$ using the SPARC plus S$^4$G samples.  A
least-squares fit is shown ($K-3.6 = -0.42\pm0.02(B-V)+0.54\pm0.01$) and symbol type
divides the sample into high and low-mass systems (divided at $10^{10.5}$).  As found
by Schombert \& McGaugh (2014), the mean $K-3.6$ color is around 0.3 with a slight
color term such that optical bluer galaxies have slightly redder $K-3.6$ colors.

Also shown in Figure \ref{bv_k36} are the SSP tracks from CG10, each track displaying
colors for ages greater than 0.5 Gyrs.  The general trend of the SSP tracks is the
same as the fitted linear slope.  Higher metallicity models have redder $K-3.6$
colors at a set $B-V$ color, and this effect can been seen in the sense that
high-mass galaxies lie above the linear fit, as expected for their higher
metallicities.  But, there is significant scatter in this trend.  For example, the 10
Gyrs models for varying [Fe/H] are, basically, a line of constant $K-3.6$ color of a
value of 0.14.  This minimizes its usefulness as a metallicity indicator, but
supports the expectation that $\Upsilon_*$ deduced from 3.6$\mu$m colors will be the
most stable under varying age and metallicity conditions and confirms the CG10
tracks.

\section{Discussion}

\subsection{Effects of Varying Stellar Population Parameters}

For our simulations, we consider a range of SFH's and mixtures of exotic stellar
population components, particularly for the low-mass end, using the Speagle \etal
shape as a baseline. We consider log SFR$_o$ = 0.0 as the crossover point from the
lower to upper main sequence, and the point where the strength of the initial SF
burst alters to produce higher mass spirals.  Table 1 displays our baseline stellar
population parameters based on the MSg and the mass-metallicity relation.  At each
log SFR$_o$ (the current SFR from the MSg), we assign a final metallicity (where all
the simulations begin with a initial [Fe/H] value of $-$1.5).  For the high-mass end,
we map log SFR$_o$ into final [Fe/H] values of $-$0.2, +0.1 and +0.2 for total
stellar masses of $10^{9.5}$, $10^{10.5}$ and $10^{11.5}$ $M_{\sun}$ respectfully.
On the low-mass end, we adopt final [Fe/H] values of $-$1.2, $-$0.7 and $-$0.4 at log
SFR$_o$ = $-$3.5, $-$2.5 and $-$1.5 respectfully.  

In addition to final metallicities, we also adopt a chemical enrichment model that
best represents the total stellar mass, gas fraction and predicted SFH.  For
low-mass, LSB-type dwarfs, with high gas fractions and low SFR over their SF
histories, we assume a slow evolution in metallicity, we adopt a linear growth in
[Fe/H] with time.  For galaxies near 10$^{10}$ $M_{\sun}$, we adopt the Prantzos
(2009) prescription.  For the high-mass end, with high initial SFR's, we assume a
faster chemical enrichment and adopt a fast form of the Prantzos (2009) prescription,
where the 80\% final enrichment is reaching in only a few Gyrs.

\begin{table}
\centering
\caption{Baseline SFH parameters}
\begin{threeparttable}
\begin{tabular}{lccc}
\hline
log SFR$_o$\tnote{a} & log $M_*$\tnote{b} & [Fe/H]$_o$\tnote{c} & [Fe/H]\tnote{d} \\
($M_{\sun}$ yr$^{-1}$) & ($M_{\sun}$) &  & rate \\
\hline
$-$3.5 & 10$^7$      & $-$1.2 & slow \\
$-$2.5 & 10$^8$      & $-$0.7 & slow \\
$-$1.5 & 10$^9$      & $-$0.4 & slow \\
$-$0.5 & 10$^{9.5}$  & $-$0.2 & normal \\
+0.5   & 10$^{10.5}$ & +0.1   & fast \\
+1.0   & 10$^{11.5}$ & +0.2   & fast \\
\hline
\end{tabular}
\begin{tablenotes}
\item[a] the current star formation rate
\item[b] total stellar mass
\item[c] final mean metallicity
\item[d] chemical evolution rate
\end{tablenotes}
\end{threeparttable}
\end{table}

The conditions outlined in Table 1 represent the baseline SFH plus chemical evolution
that fits the observations for both the high and low-mass ends of the main sequence.
We will alter then shape of the SFH (always to produce the correct slopes of the
MSg), as outlined in \S4, but not the chemical evolution scenarios.  We will also
indicate the effect of changing mean [Fe/H] in the following two color diagrams, but
maintain the same chemical enrichment for two reasons.  First, for the low-mass end,
the final metallicities are already very low and the change in the chemical growth
(and resulting metallicity distributions) are negligible at these low values.  On the
high-mass end, the rapidity of chemical enrichment under high SFR build-up results in
internal metallicity distribution that are dominated by high metallicity stars.
Thus, we found that changes in the style of chemical enrichment was negligible on the
final galaxy colors.

\begin{figure}
\centering
\includegraphics[scale=0.50,angle=0]{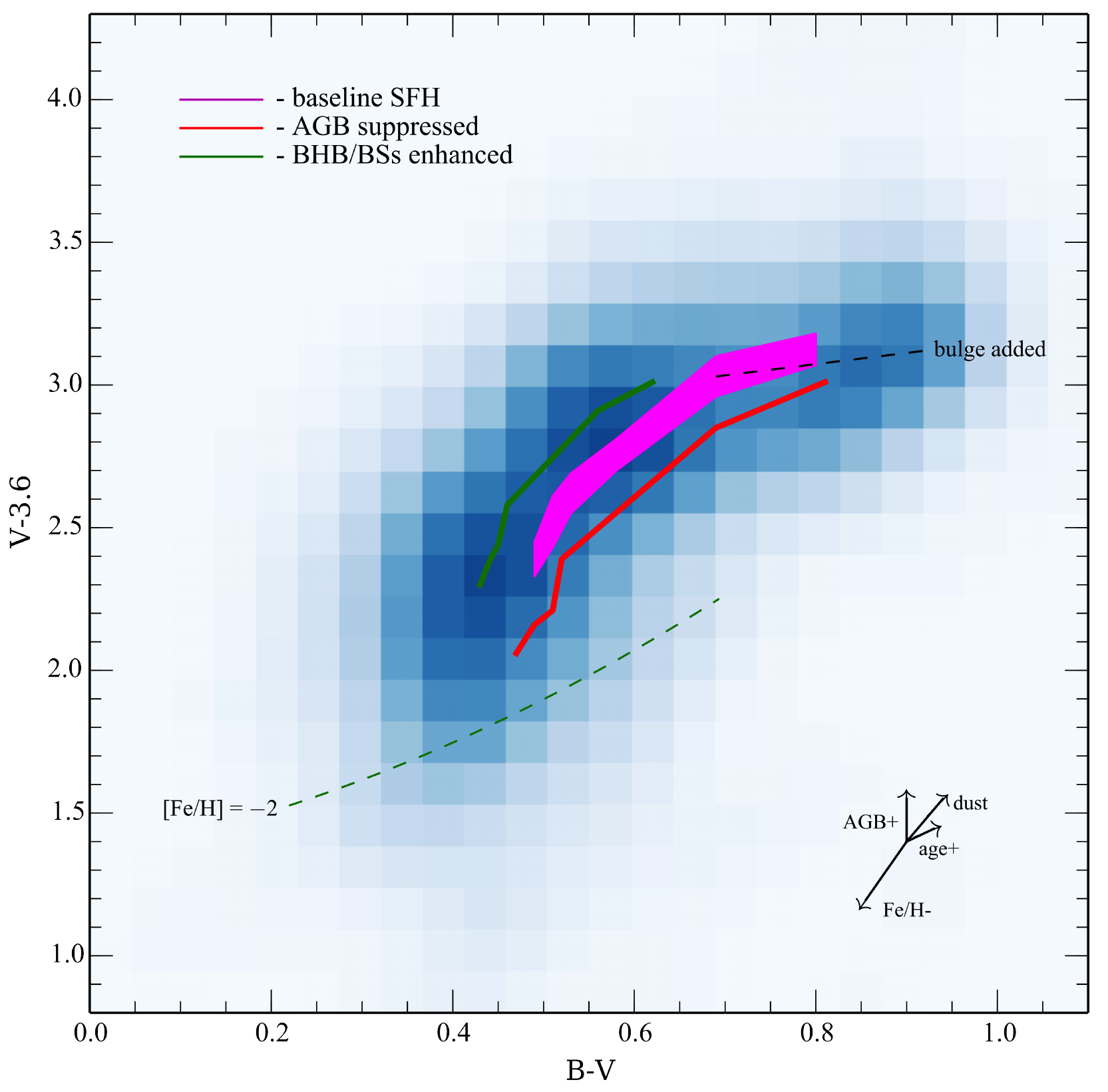}
\caption{\small The two color diagram from Figure \ref{two_color_side} drawn as a
Hess diagram where the photometric errors are used to normalize a 12x12 grid.  The
baseline SFH from Speagle \etal is shown as the magenta zone (see Table 1 for the
midline values).  The AGB
suppressed model is shown in red, the BHB+BSs enhanced model is shown in green.
Various error budget arrows are shown for changing dust, age, metallicity and AGB
component.  The dashed green line represents the lowest available metallicity SSP
model for [Fe/H]=$-$2.  The dashed black line represents the effect of a solar
metallicity disk with an increasing bulge component from 0 to 100\% to mimic the
color changes due to the morphology of early-type spirals (Sb to S0).  In general,
the baseline scenarios match the colors of high-mass star-forming galaxies, but fail
to match the low-mass galaxies by often being 
too red in optical colors and too blue in near-IR colors.
}
\label{two_color_heat}
\end{figure}

The color results for the baseline models, listed in Table 1, are shown in Figure
\ref{two_color_heat}.  The baseline model is shown as a magenta zone, using the
Speagle \etal SFH, which the width of the zone represents varying the baseline model
by 2$\sigma$ in metallicity from the mass-metallicity relation (this differs from
varying the metallicity of the stellar population, see below).  The blue line is the
BHB+BSs enhanced model, where the fraction of BHB and blue stragglers are enhanced in
agreement with the results from the CMD of LSB dwarfs (Schombert \& McGaugh 2015).
The red line represents the AGB suppressed model, also in agreement with LSB CMD's.
The dashed line displays the effects of adding an elliptical-like bulge to a solar
metallicity star forming disk in varying bulge to disk ratios (i.e., recovering large
bulge, early-type spirals).  The CG10 SSP for a [Fe/H] = $-$2 is also shown to
represent the very minimal metallicity acceptable for galaxy populations.  The
various error budget vectors are also indicated in the bottom right of the panel and
discussed below.

In general, the baseline model works well for high-mass spirals, unsurprisingly as
those galaxies on the high-mass end of the main sequence that are well mapped into
the Speagle \etal SFH.  For lower mass galaxies (with bluer colors) the color locus
is consistently too blue in $B-V$, or too red in $V-3.6$, compared to the baseline
models.  Increasing the AGB contribution serves to redden the $V-3.6$ color by the
correct amount, but is in tension with decreasing contribution by AGB's found for low
metallicity LSB dwarfs.  This shift would also make the very blue optical colors for
a majority of low-mass galaxies difficult to reconcile with redder $V-3.6$.  An
increasing blue component (BHB+BSs) is agreement with CMD diagrams of nearby dwarfs (see
Schombert \& McGaugh 2014) and matches the color locus in the sense that the low
metallicity models match the low ($B-V < 0.4$) colors of many LSB dwarfs.  Perhaps
the most realistic model is a blend of BHB+BSs enhance and AGB deficient models for
LSB dwarfs.

Our first test was to compare the effect on colors to changing forms of the IMF.
Comparison between the Salpeter, Chabrier and Kroupa IMF formulations was made for
models with [Fe/H] = $-$0.5 and solar.  The lower the metallicity, the lower the
differences in optical colors.  For example, the [Fe/H]=$-$0.5 model resulted in mean
differences of 0.001 in $B-V$.  In the near-IR, the differences were slightly higher,
at 0.005 in $V-3.6$, while relatively constant with metallicity.  From this simple
experiment, we conclude that IMF changes of these quantities were negligible with
respect to color, but may be significant for $\Upsilon_*$ calculations (see \S7).  

Next, we investigated the effects of dust on the baseline models.  We use the
standard dust model from Conroy, White \& Gunn (2010), but ignore the circumstellar
dust associated with AGB's.  We ignore AGB dust primarily because the baseline models
accurately predict the $V-K$ colors of LMC/SMC clusters without additional dust
contributions.  Any dust contamination in those clusters appears minimal, so
corrections would be inappropriate.  The error for dust in $B-V$ versus $V-3.6$ is
shown in Figure \ref{two_color_heat}.  While it has a slight metallicity dependence,
the variation is less than 20\%, and the uncertainty arrow in Figure
\ref{two_color_heat} is a median value.  Given the global dust abundance as a
function of morphological type, we expect dust to have a negligible effect on optical
colors in the bluest galaxies due to the lack of far-IR emission in those galaxy
types.

Given that the baseline models fit the high-mass end of spiral colors well, the
addition of dust (dominant at those morphological types) seems excessive.  We note
that we do correct our colors for internal extinction using the RC3 prescriptions
based on inclination.  This appears to mitigate any model dependent dust effects.
For the purposes of calculating $\Upsilon_*$ from the stellar population models, we
have ignored any additional corrections for dust other than standard corrections for
internal extinction (Sandage \& Tammann 1981).

The largest dependence in color space is, of course, metallicity.  Mostly through the
temperature of the RGB, but also significantly with respect to line blanketing in the
UV to blue portion of the spectrum.  While each of the model tracks in Figure
\ref{two_color_heat} accounts for variations in mean metallicity, and a chemical
enrichment model to produce an internal MDF, a inaccurate final assumed [Fe/H] value
will result in erroneous colors.  The error vector in Figure \ref{two_color_heat}
displays the change in color for an error of 0.5 dex in the final [Fe/H] (a fairly
extreme value).  Note that variable [Fe/H] follows the same basic slope as the color
locus, thus, the scatter in the two color diagram is not, primarily, from variations
in mean [Fe/H].

Age reflects into the error budget primarily in some error with respect to the
formation epoch.  A later formation time naturally produces bluer colors as the mean
age decreases.  While this will be addressed specifically below, by varying the SFH
model, a rough estimate of the effect of increasing age is shown in Figure
\ref{two_color_heat} for a change of 2 Gyrs in mean age.  This also roughly displays
the effect of a "frosting" model (Trager \etal 2000) where a younger stellar
population is arbitrarily added to a SFH model, perhaps simulating the merger of a
younger system.  In either case, the effects of age in the two color diagram are less
than metallicity, but notable.

\begin{figure}
\centering
\includegraphics[scale=0.50,angle=0]{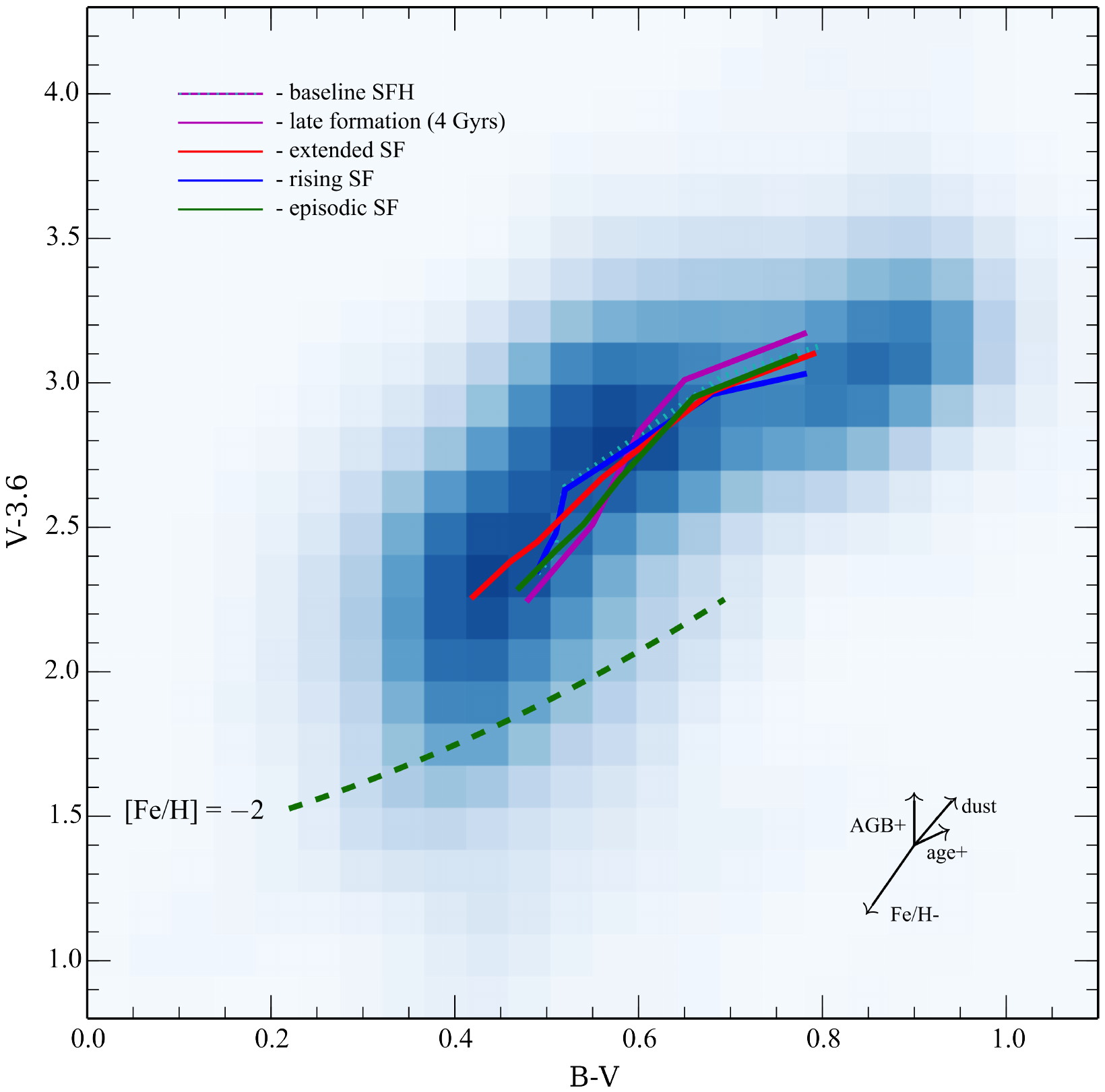}
\caption{\small Shown are the SFH models from Figure \ref{sfh_mix}.  The baseline
model from Table 1 is shown as the dotted line.  A scenario where initial star
formation is delayed by 4 Gyrs is shown as the magenta line.  A scenario where the
initial star formation burst is extended in time is shown as the red line.  A
scenario where is rising over the last 2 Gyrs (rather than declining as in the
baseline scenario) is shown as the blue line.  Lastly, an episodic scenario where the
SFR varies by 50\% over the last 2 Gyrs is shown as the green line.
}
\label{two_color_mixed}
\end{figure}

\subsection{Effects of Varying Star Formation History}

Figure \ref{two_color_mixed} displays the changes in the two color diagram for
changes in the SFH from Figure \ref{sfh_mix}.  The baseline model from Table 1 is
shown as the dotted line.  A scenario where initial star formation is delayed by 4
Gyrs is shown as the magenta line.  A scenario where the initial star formation burst
is extended in time is shown as the red line.  A scenario where the SFR is rising
over the last 2 Gyrs (rather than declining as in the baseline scenario) is shown as
the blue line.  Lastly, an episodic scenario where the SFR varies by 50\% over the
last 2 Gyrs is shown as the green line.

As can be seen in Figure \ref{two_color_mixed}, changes in the details of the SFH
scenario have only a minor effect on the colors.  The only significant change is
noted by the late SF model, which mostly extends the near-IR colors to redder colors
at higher rates of SF in recent epochs.  Aside from the late SF scenario, all the
other scenarios match the high-mass side colors.

On the low-mass side, the difficulty with all the models is reaching the blue
optical colors representative of LSB galaxies.  While many of the scenarios can
reproduce the redder near-IR colors at a particular $B-V$, none of the SFH
scenarios adequately match the colors of the blue end of every two color diagram.
Given that nearly half of the low-mass galaxies in the Cook \etal and LSB+SPARC samples
have blue UV colors, this would seem to support a rising SF scenario for a
significant fraction of the low-mass galaxies.  However, an additional blue stellar
component is required even for the rising SF scenario.

In summary, the baseline SFH scenario, outlined in Table 1, recovers the general
characteristics of the color locus from Figure \ref{two_color_side} on the high-mass
side, but predicts much redder optical colors than observed on the low-mass side.
The addition of a bulge component improves the color predictions on the high-mass
side to recover the effect of large bulges on early-type spirals.  However, the
baseline scenario does not reach the blue colors of low-mass dwarfs and LSB galaxies,
which requires some additional blue component in the optical, without a significantly
increasing in near-IR colors.  

A hot component such as BHB and BSs stars satisfies this criteria and agrees with CMD
results from nearby galaxies.  A rising SF in the last few Gyrs will also reach this
region of the two color diagrams, but a substantial rise in SF will not reproduce the
slope of the MSg (i.e., the rising SF models do not smoothly connect with the
high-mass end of the MSg).  A deficiency in AGB stars is difficult to explain in the
two color diagrams, and the deficiency found by Schombert \& McGaugh (2014) will need
to be confirmed with near-IR CMD's.

The SFH scenarios can now be mapped into predictions of $\Upsilon_*$ as a function of
galaxy color.  Before visiting the scenarios tested in the previous sections, the
issue of the IMF must be addressed.  For, while changes in the form of the IMF have
negligible effect on integrated colors, their effect on $\Upsilon_*$ is substantial.
In Figure \ref{ml_imf}, the baseline scenario is shown as a solid black line using
the Kroupa (2001) prescription for the IMF.  The dashed line displays the change in
using the Chabrier (2003) prescription.  The Kroupa prescription is more bottom heavy
in low-mass stars than the Chabrier prescription and results in an increase of about
0.04 in $\Upsilon_*$ at 3.6$\mu$m.  This represents the upper limit to error in our
$\Upsilon_*$ estimates as the exact form of the IMF in galaxies is still open to
debate (Bernardi \etal 2017).

\begin{table}
\centering
\caption{Baseline $\Upsilon_*$ fits}
\begin{threeparttable}
\begin{tabular}{lccccc}
\hline
color\tnote{a} & band & a & b & c & $\Upsilon_*$\tnote{b} \\
\hline
$B-V$   & $B$   & $-$1.187 & $+$3.480 & $-$1.522 & 1.07 \\
$B-V$   & $V$   & $-$1.224 & $+$3.120 & $-$1.271 & 1.18 \\
$V-R$   & $R$   & $-$3.331 & $+$5.827 & $-$1.788 & 1.18 \\
$V-I$   & $I$   & $+$1.157 & $-$1.160 & $+$0.147 & 1.04 \\
$V-J$   & $J$   & $+$0.763 & $-$2.268 & $+$1.570 & 0.79 \\
$V-K$   & $K$   & $+$0.657 & $-$3.118 & $+$3.417 & 0.54 \\
$V-3.6$ & $3.6$ & $+$0.933 & $-$4.932 & $+$6.123 & 0.41 \\
\hline
\end{tabular}
\begin{tablenotes}
\item[a] The stellar mass-to-light ratio in band $i$ is given by log $\Upsilon_*^i =
a(color)^2 + b(color) + c$
\item[b] The mass-to-light value for a solar metallicity
galaxy or 10$^{10}$ $M_{\sun}$ on the main sequence
\end{tablenotes}
\end{threeparttable}
\end{table}

\subsection{Deduced Color-$\Upsilon_*$ Relationships}

Several studies have focused on the $Spitzer$ wavelengths to explore the range in
$\Upsilon_*$ with galaxy color.  Eskew \etal (2012) find a mean $\Upsilon_*$ of 0.57
at 3.6$\mu$m from a study of LMC star clusters.  Meidt \etal (2014) correlate 3.3-4.5
color with population models for a mean $\Upsilon_*$ of 0.6 at 3.6$\mu$m.  Querejeta
\etal (2015), using S$^4$G data, finds a similar color term to previous studies with
a zeropoint of $\Upsilon_*$=0.48.  Common to all these studies is a very small
dependent on color and a nearly dust free estimate of $\Upsilon_*$.  All these model
values bracket the values found for our baseline model with the Kroupa IMF.

On the optical side, Taylor \etal (2011) find an empirical relationship between model
deduced $\Upsilon_*$ and SDSS colors $g-i$ (their equation 7).  Their $\Upsilon_*$
versus color relationship produces a value of 0.67 in the $i$ band for the bluest
galaxies, rising to 1.35 for the reddest in our sample.  This exactly matches the
predictions from our baseline models (0.65 to 1.33) even though they use a Chabrier
IMF (see also Lopez-Sanjuan \etal 2018).

Our reddest models are consistent with the bluest ellipticals from Cappellari \etal
(2006) using $\Upsilon_*$ based on SAURON dynamical estimates.  These values are also
consistent with single burst models from our own study of ellipticals (Schombert
2016).  Conroy \& van Dokkum (2012) find the $\Upsilon_*$ at $K$ increases with
galaxy mass from a value of 1.0 for low-mass ellipticals to a maximum of 1.5 for
high-mass ellipticals.  Given $B-V$ colors of 1 for bright ellipticals, these values
would match an extension of our Kroupa IMF models to redder colors.  However,
ellipticals are an extension in color space that, presumingly, have very different
SFH's (e.g., a large initial burst).  In addition, Conroy \& van Dokkum (2012) find
that the IMF in ellipticals becomes more bottom heavy as SF timescales become shorter
(i.e., more massive ellipticals, Thomas \etal 2005, see a dissenting view in Parikh
\etal 2018).  This appears to uncover some underlying physics in star formation where
more intense star formation events result in the production of a greater number of
low-mass stars.

\begin{figure*}
\centering
\includegraphics[scale=0.80,angle=0]{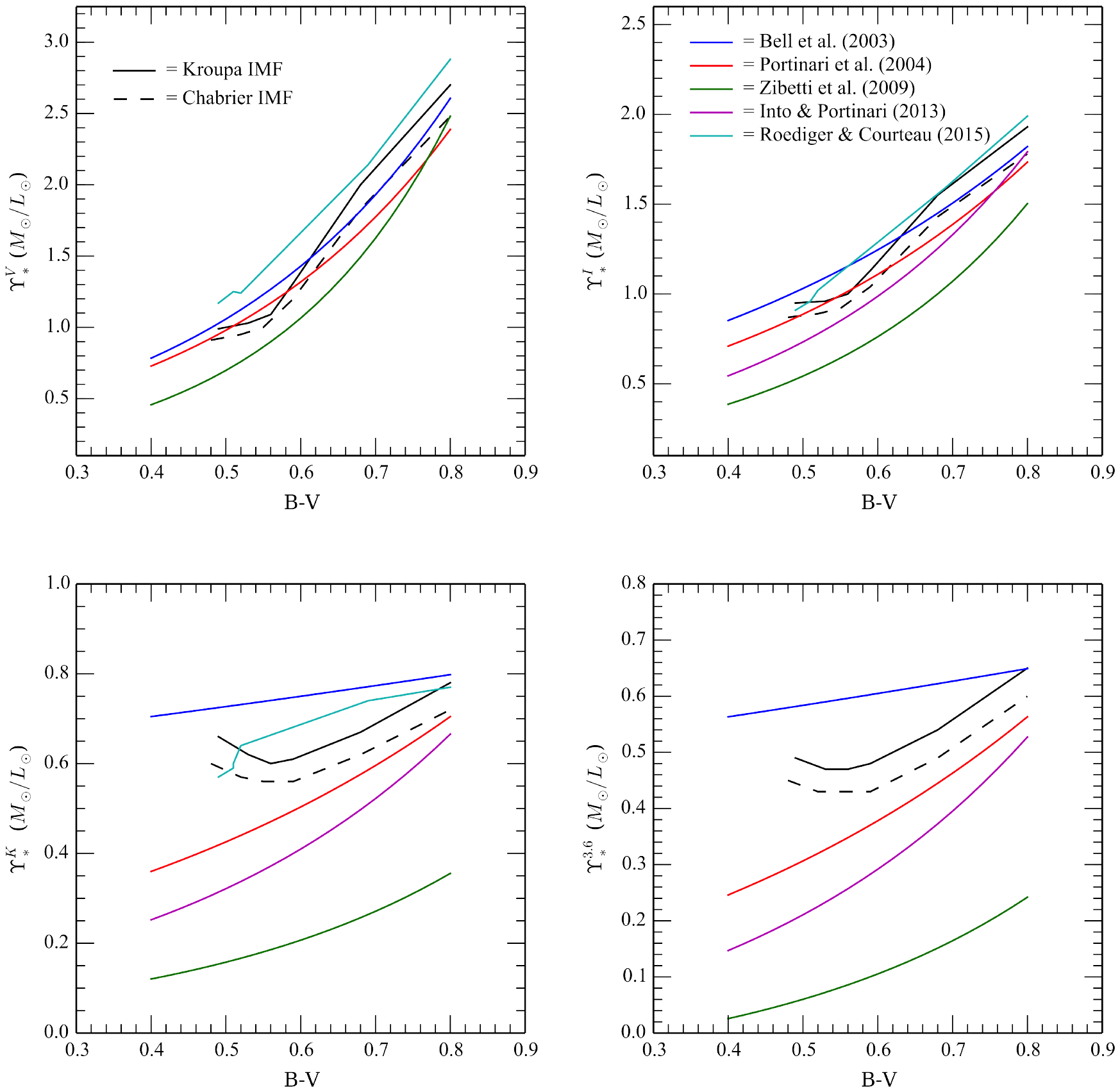}
\caption{\small The $\Upsilon_*$ versus color relations for four bandpasses
($V$, $I$, $K$ and 3.6$\mu$m).  The solid black line displays the results for the
baseline model from Table 1 using a Kroupa IMF, dashed line is for a Chabrier IMF. Five
models from the literature as also shown (Bell \etal 2003, Portinari \etal 2004,
Zibetti \etal 2009, Into \& Portinari 2013 and Roediger \& Courteau 2015).  Our
current models agree in the optical bandpasses, but there is significant range in the
near-IR bandpasses as outlined in McGaugh \& Schombert (2014).
}
\label{ml_imf}
\end{figure*}

Armed with the change in behavior in the IMF for ellipticals, we can attempt to
extrapolate these conditions to the SFH of gas-rich spirals and dwarfs.  By
definition, the SF in low-mass spirals and dwarfs is extended in time and does not
proceed in strong bursts as in ellipticals.  While it is expected to be episodes of
stronger SF based on CMD studies of nearby dwarfs (Dalcanton \etal 2009, McQuinn
\etal 2010), none should reach the intensity of an elliptical burst.  Our initial
estimates indicate that the IMF should be bottom light for low-mass star-forming
galaxies, increasing slightly in the number of low-mass stars as we get to higher
mass spirals with stronger past SF.  With respect to our baseline model, we would
estimate that $\Upsilon_*$ at 3.6$\mu$m would be 0.45 for the bluest galaxies (see
Figure \ref{ml_imf}) rising to values of 0.5 at intermediate colors and 0.6 for the
highest mass spirals.

Figure \ref{ml_imf} summarizes our baseline model comparison to other models in the
literature from optical to near-IR bandpasses.  The Bell \etal (2003), Portinari
\etal (2004), Zibetti \etal (2009) and Into \& Portinari (2013) studies were
discussed in McGaugh \& Schombert (2014).  They all use a variety of IMF's and AGB
prescriptions (outlined in McGaugh \& Schombert 2014, Table 5).  New to our
comparisons is the study by Roediger \& Courteau (2015) which uses the newer
population models also used in this study.  The difference in the optical
$\Upsilon_*$ values are primarily due to different assumptions in the SFH of the
models.  Despite the variance in SFH assumptions, the $\Upsilon_*$ values are very
similar and follow the same trends with color.  The near-IR $\Upsilon_*$ models
diverge significantly, primarily on changes to the treatment of AGB's and updated
isochrones.  Roediger \& Courteau use the same isochrones as this study and produce
similar $K$ values (they did not investigate $Spitzer$ bandpasses).

\begin{figure*}
\centering
\includegraphics[scale=0.80,angle=0]{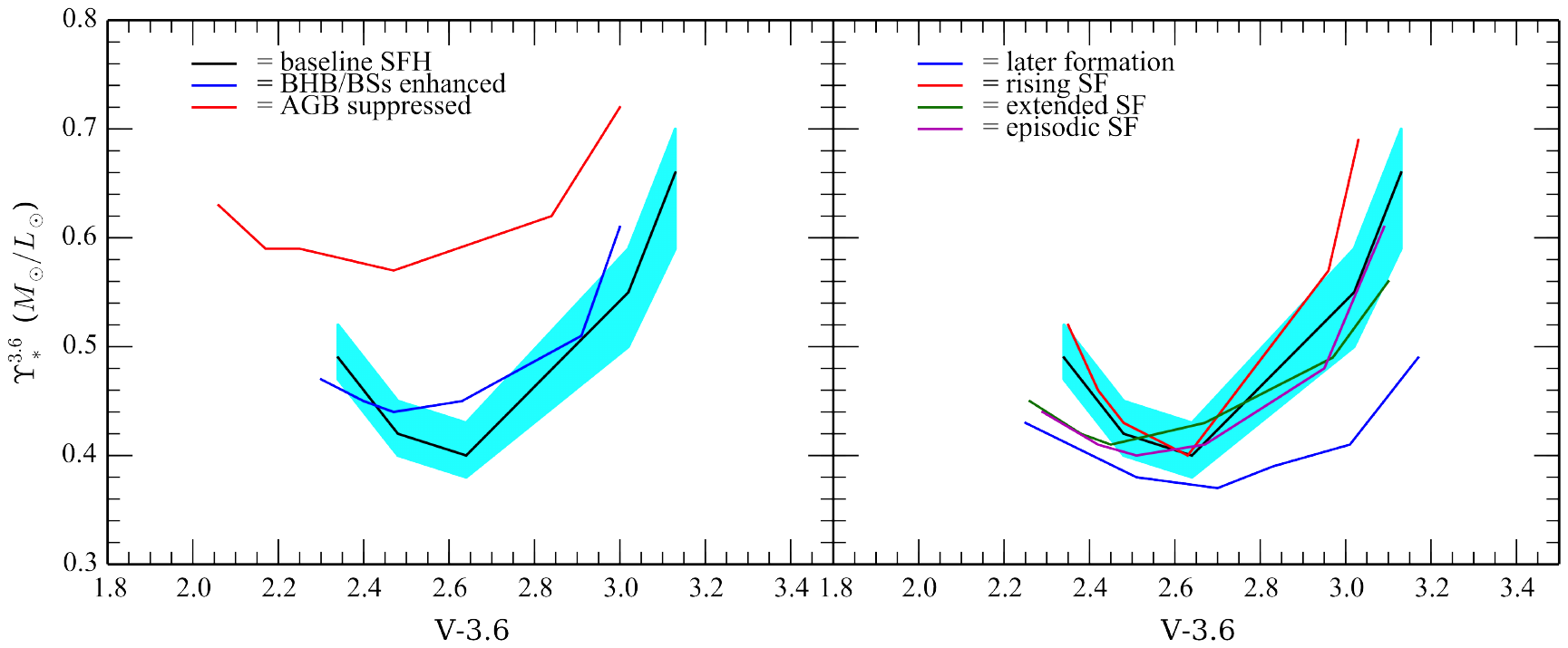}
\caption{\small A comparison of the effects of different stellar population
parameters and SFH's on $\Upsilon_*$ at 3.6$\mu$m.  The various models are discussed
in the text.  While most studies predict similar $\Upsilon_*$ values in the optical
bandpasses, they all diverge at near-IR bandpasses due to varying prescriptions on
later SFH effects and AGB treatment.  The shade area defines a region for models where the
final SFR and final [Fe/H] values were varied by 2$\sigma$ from their deduced values
in the MSg and mass-metallicity relations.
}
\label{ml_side}
\end{figure*}

The variations in model assumptions outlined in Figures \ref{two_color_heat} and
\ref{two_color_mixed} are presented in how they effect $\Upsilon_*$ in Figure
\ref{ml_side}.  The baseline model, as outlined in Table 1, are displayed as the
solid black line.  In addition, the baseline models were expanded to consider
different values for final SFR and [Fe/H] where at a particular stellar mass we vary
the SFR and [Fe/H] to cover a 2$\sigma$ change in the MSg and mass-metallicity
relations.  The change in color is primarily driven by metallicity, so these
dispersion models track the [Fe/H] vector in color space, but can be significant in
$\Upsilon_*$ estimates as shown by the colored regions in Figure \ref{ml_side}.  Second
order polynomial fits to the baseline models, across all the colors in the sample, are
listed in Table 2.  Also found in Table 2 is the mean value for a solar metallicity galaxy
(i.e., one at the turnoff point in the MSg, or $10^{10}$ $M_{\sun}$).

Changes in the assumed stellar population parameters, such as the fraction of BHB,
BSs and AGB stars have the obvious effects.  BHB and BSs stars, which have a
significant effect on optical colors, but do not have a noticeable change on the
$\Upsilon_*$ values at 3.6$\mu$m.  The absence of a significant fraction of AGB stars
has a large effect on $\Upsilon_*$ as the 3.6$\mu$m where luminosity decreases by
20\%.  As this AGB absence is noted in CMD diagrams of nearby LSB dwarfs, this is a
concern for accurate $\Upsilon_*$ estimates for low SFR dwarfs (the high SFR dwarfs,
such as the ANGST sample, do not display this deficiency).

The right panel in Figure \ref{ml_side} displays variation in the assumed SFH as
discussed in \S4.  All the considered scenarios result in lower $\Upsilon_*$ values
compared to the baseline scenario.  The later and extended SF scenarios have lower
mean  $\Upsilon_*$ (of about 0.1 dex) due to the fact that both these scenarios input
higher fractions of younger stars compared to the other scenarios.  Decreasing the
SFR in the past would move the mean  $\Upsilon_*$ upward, but this would place these
scenarios in tension with the main sequence results, in that the total star formation
rate needs to be lower to reproduce the correct final stellar masses which,
typically, alters the predicted slope of the MSg.  As the slope of the MSg is the
least well defined aspect to star-forming galaxies, there is some leeway in possible
star formation histories.  

If the deficiency in AGB stars for LSB dwarf is confirmed, then the decrease in
$\Upsilon_*$ by later or extended SFH's is offset by the bottom heavy nature of
AGB-light stellar populations.  This may explain the conclusion from Schombert \&
McGaugh (2014) on the surprising stability of $\Upsilon_*$ as a free parameter in
fits to the radial acceleration relation as population effects are balanced by SFH
scenario changes for lower mass galaxies.

Lastly, we note, for accurate mass models, population gradients can play a role.  In
particular, the need to adjust $\Upsilon_*$ for a bulge component.  As noted in
Figure \ref{two_color_heat}, the addition of a bulge component to a solar metallicity
model, by B/D ratio to colors, exactly matches the red end of the two color diagram.
The effect on $\Upsilon_*$ would be effective raising the value of the reddest
galaxies from 0.6 to 0.7 for the early-type spirals, and 0.8 for S0 galaxies.

\subsection{Uncertainties in $\Upsilon_*$ Models}

While having $\Upsilon_*$ to color relationships, such as Figure \ref{ml_side}, are
the ultimate goal for obtaining stellar mass from a luminosity value, the correct
application of this value requires knowledge of their accuracy and uncertainty.
Accuracy, in this context, refers to the effect that observational error (in this
case, the error in the observed galaxy color) has on the $\Upsilon_*$ value.
Uncertainty refers to the range in $\Upsilon_*$ values for a reasonable range in
model parameters for the particular galaxy color or mass.

Accuracy is relatively easy to calculate.  For a known error in color, one searches for
the two models that satisfy the upper and low bounds in color (i.e., the $\Upsilon_*$
values for two $V-3.6$ colors in Figure \ref{ml_side}).  This range will be higher
for bluer colors (e.g., $B-V$ in Figure \ref{ml_imf}), which is why near-IR colors
are preferred for $\Upsilon_*$ work.  For typical near-IR color errors in the 0.05
range, the resulting errors in $\Upsilon_*$ (at 3.6) were 0.04.

\begin{figure}
\centering
\includegraphics[scale=0.45,angle=0]{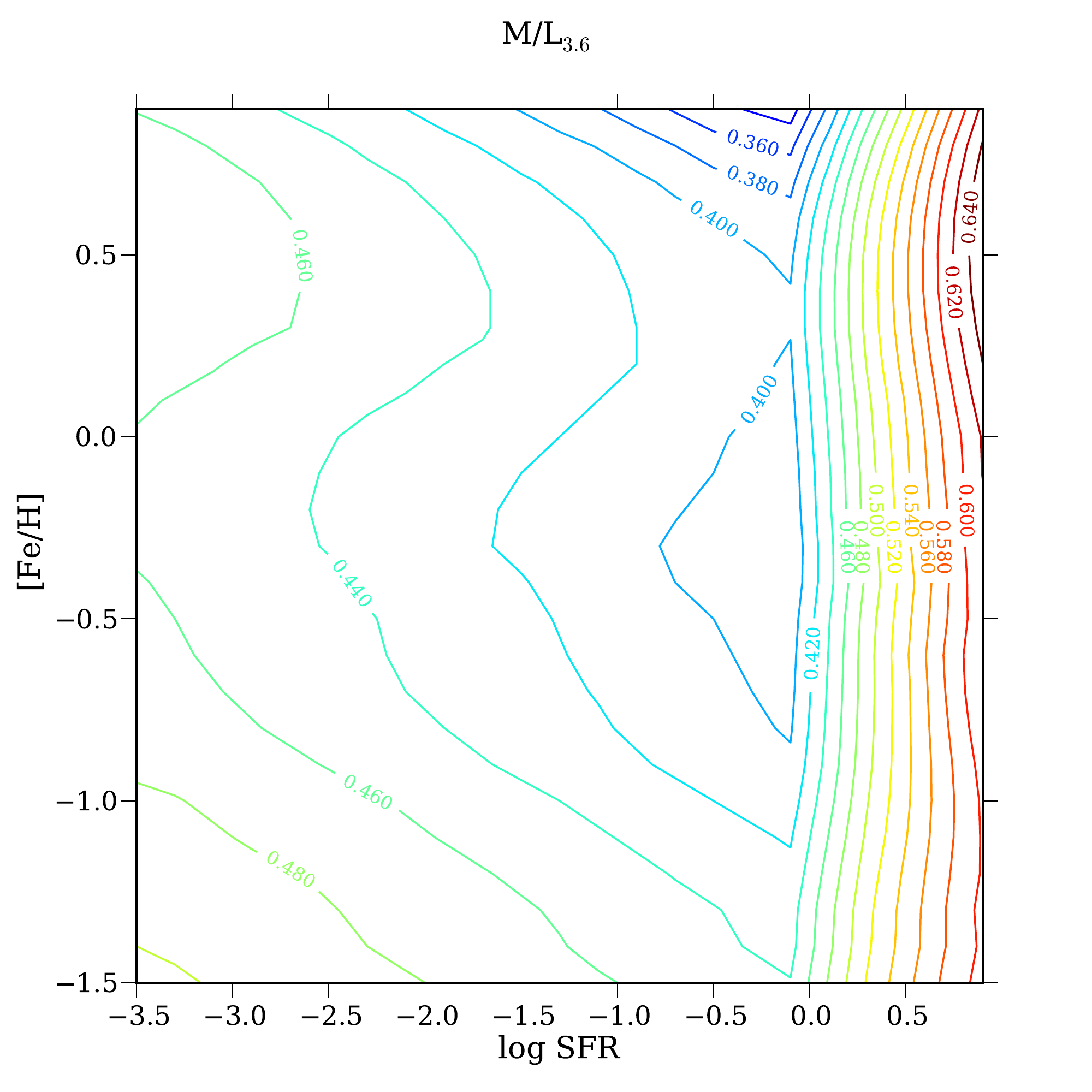}
\caption{\small The range of calculated $\Upsilon_*$ values for models with a range
of current SFR and metallicities (the two defining model inputs where SFR defines the
SFH shape, and therefore the ages of the underlying stellar population, and current metallicity
which defines the chemical enrichment mode, and therefore the metallicity
distribution in the underlying stellar population).  The more compressed contours
above log SFR = 0 reflects the shallower slope to the high-mass MSg, which drives an
older stellar populations from stronger and earlier star formation.  Below log SFR =
0, the models and deduced $\Upsilon_*$ values are robust to small errors in the MSg
and mass-metallicity relations (see text).
}
\label{ml_contour}
\end{figure}

Uncertainty questions the validity of a particular model to a particular galaxy
color or mass.  This, unsurprisingly, has a larger impact on $\Upsilon_*$ values than
color errors.  Uncertainty in the models reflects either error in the choice of the
input parameters (in our scenarios, current SFR and metallicity) or inappropriate choice
of stellar population parameters (e.g., an enhanced BHB scenarios).  The
latter, choice of population parameters, results in a discrete shift in $\Upsilon_*$ values
(see Figure \ref{ml_imf}) and can not quantified as error on $\Upsilon_*$.
However, error in the input parameters does reflect a inherent unknown from the
scatter in the MSg plus mass-metallicity relation.  Dispersion in the measured
current SFR produces a range in deduced SFH.  Range in assumed metallicity alters the
chemical enrichment model.  Both can have significant effects on the final deduced
galaxy colors.

To access the impact on the scatter in the MSg and mass-metallicity relation, we can
consult a $\Upsilon_*$ contour plot such as shown in Figure \ref{ml_contour}.  Here a
range of SFR and [Fe/H] values are plotted with their resulting $\Upsilon_*$ values.
The SFR value contains two uncertainties, the scatter in the MSg relation in the SFR
direction plus the scatter in total stellar mass (since the integrated SFH from the
SFR value produces the total stellar mass).  However, the total stellar mass error is
small (small changes in $\Upsilon_*$ with SFR) and can be iterated.

As noted in \S2, the scatter in the MSg is around 0.4 dex on the low-mass end and
around 0.1 on the high-mass end.  While the scatter in the mass-metallicity relation
is around 0.3 dex (increasing to lower galaxy masses).  For particular regions in
Figure \ref{ml_contour} this can mean the model $\Upsilon_*$ values are robust (e.g.,
below log SFR = $-$1) or increasingly sensitive to the assumed SFR value (e.g., above
log SFR = 0).

\section{Summary}

The premise of this study is that knowledge of the SFH in star-forming galaxies as
given by the main sequence relation, combined with present-day colors and flexible
stellar population models, allows for an understanding of the underlying stellar
population in terms of their total mass and luminosity.  Previous studies have
indicated that the mass-to-light ratio ($\Upsilon_*$) varies with the rate of star
formation, but that this maps smoothly into galaxy color, which is driven by the same
population changes that reflect into $\Upsilon_*$.

Completely accurate knowledge of the distribution of age and metallicity of stars in
star-forming galaxies will be an elusive goal.  However, the numerical experiments in
\S5 and \S6 indicate the range of possible SFH's is fairly narrow plus results in a
similar distributions of the various types of stars that dominate galaxy colors.
This narrow range in stellar populations also reflects into well defined color-$\Upsilon_*$
relations from which an accurate mass-to-light ratio can be extracted across most
optical and near-IR bandpasses.

Our conclusions can be summarized as follows:

\begin{itemize}

\item The main sequence for star-forming galaxies divides into two parts (see Figure
\ref{main_seq}); 1) the high-mass end (weary giants, $M_* > 10^{10.5} M_{\sun}$)
which has been investigated with redshift and has a shallow slope (Speagle \etal
2014), and 2) the low-mass end (thriving dwarfs, $M_* < 10^{10.5} M_{\sun}$),
explored by McGaugh, Schombert \& Lelli (2017) and Cook \etal (2014), which has a
steeper slope, parallel to a line of constant SF over time and is sub divided by
$FUV-NUV$ color.  The division by UV colors across the line of constant SF signals a
separation between galaxies with declining versus rising SFR over that last Gyr.

\item The various color locus for star-forming galaxies has coherence across
bandpasses (i.e., red galaxies are red in all colors), but has a great deal of
scatter in excess of the observational error suggesting a wide variance in age and
metallicity.  The fact that color divides nominally by galaxy stellar mass (see Figure
\ref{two_color_side}) implies that metallicity is the strongest driver for color with
increasing importance to recent SF at the low-mass end.

\item The SFH of the high-mass end of the MSg is well-defined by the study of 
Speagle \etal (2014).  This can be qualified, roughly, as a moderate,
wide initial burst of SF with a shallow decline to the present epoch.  To reproduce
the high-mass end of the MSg, this decline must be steeper with increasing final
stellar mass.  In adopting a similar SFH shape for the low-mass end of the MSg, one
must have a slightly longer initial burst with nearly constant SFR at the inflection
point of $10^{10.} M_{\sun}$ with slightly declining SFH with lower mass (see Figure
\ref{speagle_sfh}).  These are the baseline SFH used in Table 1; however, a range of
different SFH can produce the same slope on the low-mass end of the MSg (see Figure
\ref{sfh_mix}) but produce different color and metallicity relations.

\item Stellar population models mix age (the assumed SFH) and metallicity (the
assumed chemical enrichment model) to produce a composite stellar population that
generates the observed integrated luminosity and colors.  Of the many components to a
stellar population, one is critical to $\Upsilon_*$ (the IMF) and two are critical to
galaxy colors (AGB treatment and the BHB+BSs component).  With respect to the
low-mass end of the MSg and near-IR colors, the AGB component, while matching LMC/SMC
cluster colors, appears deficient in some LSB galaxies (Schombert \& McGaugh 2015)
and enhanced in other star-forming dwarfs (Dalcanton \etal 2009).  In a similar
fashion, an enhanced BHB+BSs population (expected at low metallicities) alters,
significantly, the optical colors.  In order to match the galaxy color locus (Figure
\ref{two_color_heat}), baseline stellar populations are adequate, but enhanced AGB
and/or enhanced BHB+BSs components are required at the low-mass end.

\item Variations in SFH scenarios, within limits to maintain the slope and zeropoint
of the MSg, do not produce large variations in galaxy colors (see Figure
\ref{two_color_mixed}).  Recent SF can drive optical colors to very blue values ($B-V
< 0.4$), but changes in near-IR colors are difficult to reproduce with different SFH
scenarios.

\item Existing $\Upsilon_*$ versus color relationships in the literature converge in
their predictions in optical bandpasses (McGaugh \& Schombert 2014), but are wildy
discrepant in the near-IR bandpasses (see Figure \ref{ml_imf}).  Recent models (such
as Roediger \& Courteau 2015) offer a more coherent view in the near-IR due to
improved treatment of intermediate aged stars.  Our baseline models predict nearly
constant $\Upsilon_*$ with color in the near-IR, with the most significant effect
being the choice of the IMF (with a variation of 0.1 dex between bottom heavy versus
bottom light scenarios).

\item Blending stellar population components, IMF effects and SFH scenarios produces
the estimates displays in Figure \ref{ml_side}, where $\Upsilon_*$ can be estimated
with a single optical to near-IR color.  The dispersion on $\Upsilon_*$ can be
estimated as a function of stellar population effects (left panel) versus SFH effects
(right panel).  The color region in Figure \ref{ml_side} provides some guidance for a
nominal change in the input parameters as given by the uncertainty in the MSg and the
mass-metallicity relations.  Aside from radical departures from estimates of the
contribution from AGB stars or extremely later epoch's of initial star formation, all
the considered scenarios fall within the baseline model's zone of variance.  A
simplistic, first-order description for the baseline model is constant $\Upsilon_*[3.6]$
(approximate 0.45) until a galaxy color of $V-3.6 = 2.8$ followed by a linear rise to
elliptical value of 0.7 by the reddest colors of 3.2.  These models and a
$\Upsilon_*$ calculator is available at the SPARC website.

\end{itemize}

The model determination of $\Upsilon_*$ at various bandpasses is critical to many low
and high redshift studies of stellar mass in galaxies.  A well studied model from the
UV to near-IR allows for an application to high redshift systems which will account for
redshifted filters as well as different epochs being sampled with cosmic time.  The
zero redshift models are important as applied to the baryon Tully-Fisher relation
(McGaugh 2012) and the radial acceleration relation (McGaugh, Lelli \& Schombert
2016).  These models form the core of our future studies to convert galaxy photometry
into stellar masses to study the relation between baryons and dark matter.

\section*{Acknowledgements}
Software for this project was developed under NASA's AIRS and ADP
Programs. This work is based in part on observations made with the Spitzer Space
Telescope, which is operated by the Jet Propulsion Laboratory, California Institute
of Technology under a contract with NASA.  Support for this work was provided by NASA
through an award issued by JPL/Caltech. Other aspects of this work were supported in
part by NASA ADAP grant NNX11AF89G and NSF grant AST 0908370. As usual, this research has
made use of the NASA/IPAC Extragalactic Database (NED) which is operated by the Jet
Propulsion Laboratory, California Institute of Technology, under contract with
the National Aeronautics and Space Administration.

\appendix
\section{$\Upsilon_*$ Webtool}

The range of metallicities and different population parameters makes a tabulated
presentation of the models impossible.  Instead, a webtool is offered the community
(http://abyss.uoregon.edu/$\sim$js/sfh) to provide either 1) colors and $\Upsilon_*$
values per galaxy stellar mass or 2) a user selected $V-3.6$ color value and error.

The first option has two paths.  First, the user can enter either a SFR, galaxy
stellar mass ($M_*$) or final metallicity ([Fe/H]).  The webtool will then convert
any of the three parameters into the other two using the MSg and mass-metallicity
relation discussed in the text (see Table 1).  The total stellar mass determines
the rate of chemical evolution (slow, normal or fast, see \S5.1).  The selected SFR
determines the SFH input to the models.  The selected final [Fe/H] value plus
enrichment model determines the metallicity at any particular epoch.  The user can
override the values deduced from the main sequence and mass-metallicity relations by
selecting their own SFR, [Fe/H] and chemical enrichment values, although the physical
validity of many possible combinations is questionable.

A second option is to enter a $V-3.6$ color and uncertainty in that color.  The
webtool will present the best fit model for that color, and the range in $\Upsilon_*$
values due to the uncertainty in the color.  This always assumes the standard SFR and
[Fe/H] values outlined in the text (see Table 1).

Lastly, the user can adjust any of the above models for various enhancements in the
underlying stellar populations (e.g., AGB, BHB, BSs populations).  These models will
continue to apply standard SFR and [Fe/H] values per stellar mass, but use altered
stellar population parameters discussed in \S5.  Again, the meaning and value of many
of these extreme models is up to the user to decide.

\end{document}